\documentclass[fleqn,usenatbib]{mnras}

\usepackage{newtxtext,newtxmath}
\usepackage[T1]{fontenc}

\DeclareRobustCommand{\VAN}[3]{#2}
\let\VANthebibliography\thebibliography
\def\thebibliography{\DeclareRobustCommand{\VAN}[3]{##3}\VANthebibliography}

\usepackage{graphicx}	% Including figure files
\usepackage{amsmath}	% Advanced maths commands

\newcommand{\jr}{\color{black}}

\title[Under the light of a new star]{Under the light of a new star: evolution of planetary atmospheres through protoplanetary disc dispersal and boil-off}

\author[Rogers,  Owen \& Schlichting]{
James G. Rogers$^{1}$\thanks{E-mail: jamesrogers@ucla.edu}, James E. Owen$^{2}$ \& Hilke E. Schlichting$^{1}$
\\
$^{1}$Department of Earth, Planetary, and Space Sciences, University of California, Los Angeles, 595 Charles E. Young Drive East, Los Angeles, CA 90095, USA\\
$^{2}$Astrophysics Group, Department of Physics, Imperial College London, Prince Consort Rd, London, SW7 2AZ, UK\\
}

\date{Accepted XXX. Received YYY; in original form ZZZ}

\pubyear{2023}

\begin{document}
\label{firstpage}
\pagerange{\pageref{firstpage}--\pageref{lastpage}}
\maketitle

% Abstract of the paper
\begin{abstract}
The atmospheres of small, close-in exoplanets are vulnerable to rapid mass-loss during protoplanetary disc dispersal via a process referred to as `boil-off', in which confining pressure from the local gas disc reduces, inducing atmospheric loss and contraction. We construct self-consistent models of planet evolution during gaseous core accretion and boil-off. As the surrounding disc gas dissipates, we find that planets lose mass via subsonic breeze outflows which allow causal contact to exist between disc and planet. Planets initially accrete of order $\sim 10\%$ in atmospheric mass, however, boil-off can remove $\gtrsim 90\%$ of this mass during disc dispersal. We show that a planet's final atmospheric mass fraction is strongly dictated by the ratio of cooling timescale to disc dispersal timescale, as well as the planet's core mass and equilibrium temperature. With contributions from core cooling and radioactivity, we show that core luminosity eventually leads to the transition from boil-off to core-powered mass-loss. We find that smaller mass planets closest to their host star may have their atmospheres completely stripped through a combination of boil-off and core-powered mass-loss during disc dispersal, implying the existence of a population-level radius gap emerging as the disc disperses. We additionally consider the transition from boil-off/core-powered mass-loss to X-ray/EUV (XUV) photoevaporation by considering the penetration of stellar XUV photons below the planet's sonic surface. Finally, we show that planets may open gaps in their protoplanetary discs during the late stages of boil-off, which may enhance mass-loss rates. 
\end{abstract}


%%%%%%%%%%%%%%%%%%%%%%%%%%%%%%%%%%%%%%%%%%%%%%%%%%

%%%%%%%%%%%%%%%%% BODY OF PAPER %%%%%%%%%%%%%%%%%%

\section{Introduction} \label{sec:intro}
The evolution of protoplanetary discs is fundamentally linked to planet formation and evolution. For the observed population of low-mass, close-in exoplanets \citep[e.g.][]{Howard2012,Fressin2013,Silburt2015,Mulders2018,Zink2019}, it is believed that many host primordial H/He dominated atmospheres \citep[e.g.][]{Weiss2014,JontofHutter2016,Benneke2019,Rogers2023_waterworld} that were likely accreted while immersed in their nascent disc \citep[e.g.][]{Rafikov2006,Lee2014,Piso2014,Ginzburg2016}. Once the disc has dispersed, atmospheric escape, driven by stellar irradiation such as photoevaporative \citep[e.g.][]{Owen2013,LopezFortney2013,Jin2014,Howe2015,ChenRogers2016} or core-powered mass-loss \citep[e.g.][]{Ginzburg2018,Gupta2019}, is typically accredited to the eventual carving of the observed paucity of planets inside the `radius gap' \citep[e.g.][]{Fulton2017,VanEylen2018,Berger2020b,Petigura2022,Ho2023}. 

To correctly model the consequences of photoevaporation and core-powered mass-loss on the small exoplanet population, one must carefully consider their initial conditions. It is likely that these processes are not the first time that small close-in exoplanets are at risk of losing atmospheric mass. Whilst protoplanetary discs survive for $\sim 3-10$ Myr, the inner regions ($\lesssim 1\text{ AU}$) have been observed to disperse on timescales of $\sim 10^5$ yrs \citep[e.g.][]{Kenyon1995,Ercolano2011,Koepferl2013}. This rapid disc dispersal has strong implications for the atmospheric evolution of close-in exoplanets since the sudden reduction in the disc gas' confining pressure results in dramatic atmospheric escape \citep{Ikomi2012}. This process, in which a planet is illuminated by its young star for the first time, is referred to as `boil-off' \citep{Owen2016} or `spontaneous mass-loss' \citep{Ginzburg2016,Misener2021}. 

{\jr{Population-level evidence for boil-off was hinted at in \citet{Rogers2023}, in which \textit{Kepler}, \textit{K2} and \textit{TESS} planets with well-constrained masses and radii were used to statistically infer the relation between planetary core masses and initial atmospheric masses at the end of the disc lifetime. Their results were found to be consistent with the predictions from \citet{Ginzburg2016}, in which analytic arguments had been used to understand the atmospheric evolution of small planets prior and post-disc dispersal. \citet{Ginzburg2016} showed that boil-off will likely set the initial conditions for late-time escape mechanisms such as photoevaporation and core-powered mass-loss, which are thought to control small exoplanet demographics. It is, therefore, imperative to better understand the governing physics at play during boil-off.}}

% {\jr{Besides boil-off, other possible solutions to this discrepancy do exist. As discussed, the inclusion of dust opacities will reduce the cooling (and therefore accretion) rate of planet atmospheres \citep[e.g.][]{Lee2014,Lee2015}. In addition, forming the planets at the very end of the disc lifetime \citep[e.g.][]{Ikomi2012,LeeChiang2016,Lee2018} has also been shown to reduce the amount of accreted gas. One can also adapt formation models to include atmospheric loss during giant mergers \citep{Liu2015,Inamdar2016,Ormel2021}, which would result in potentially significant atmospheric mass loss and the production of heat, which would take typically kyrs to disperse \citep{Biersteker2019}. Results from 3D simulations have also shown that the recycling of high-entropy gas during the accretion phase can act to reduce the final atmospheric mass of a planet by preventing cooling, although we note that the efficacy of this mechanism is currently under debate \citep{Kurokawa2018,Ormel2015,Fung2015,Cimerman2017,AliDib2020,Chen2020,Bailey2023,Savignac2023}.}}

In this paper, we investigate the effects of boil-off on the exoplanet population by explicitly modelling the transition from gas accretion to atmospheric escape, for a planet embedded in a dissipating gas disc. Previously, boil-off was modelled in \citet{Ikomi2012} under the assumption that planet atmospheres were in hydrostatic equilibrium with the surrounding disc. This approach typically resulted in planets losing their entire atmosphere during disc dispersal since the only way to remain in hydrostatic equilibrium with a vanishing disc is to host a vanishing atmosphere. At the point of disc dispersal $t_\text{disp} \sim 3-10$ Myrs, the cooling timescale for a planet's atmosphere is likely to have an upper limit of $\tau_\text{cool} \lesssim t_\text{disp}$ since cooling timescales approximately track a planet's age and planets can form at any point during the disc's lifetime. Recall, however, that typical inner disc dispersal timescales are $\tau_\text{disp} \sim 10^5$ yrs; thus, a planet's cooling timescale $\tau_\text{cool}$ is likely greater than the disc dispersal timescale $\tau_\text{disp}$. If this is the case, a planet may be unable to thermally respond to any changes in external pressure due to disc dispersal. It will undergo adiabatic expansion on timescales of order $\tau_\text{disp}$. External heating (from the star and/or disc) will heat the expanding material, causing atmospheric escape via a rapid hydrodynamic outflow. 

In \citet{Owen2016}, the case of non-hydrostatic equilibrium was explored with numerical evolutionary models. Based on the prior timescale argument, they assumed the disc dispersed instantaneously and material was removed from the planet via an isothermal Parker wind \citep{Parker1958}. They found that up to $\sim 90\%$ of atmospheric mass was removed on timescales of $\sim 10^5$ yrs; however, this was based on initial conditions in which the planet's photospheric radius $R_\text{ph}$ was equal to its Bondi radius $R_\text{B}$:
\begin{equation}
    R_\text{B} = \frac{GM_\text{p}}{2 c_\text{s}^2},
\end{equation}
where $M_\text{p}$ is the planet mass, {\jr{$G$ is the gravitational constant and $c_\text{s}$ is the isothermal sound speed, defined as:
\begin{equation} \label{eq:soundspeed}
    c_\text{s} = \sqrt{ \frac{k_\text{B} T_\text{eq}}{\mu m_\text{H}}},
\end{equation}
where $k_\text{B}$ is the Boltzmann constant, $T_\text{eq}$ is the planet's equilibrium temperature, $m_\text{H}$ is the mass of a hydrogen atom, and we use $\mu=2.35$ \citep[e.g.][]{Anders1989}.}} The aforementioned assumption of $R_\text{ph} = R_\text{B}$ meant that the planet was initially extremely inflated and fully convective, i.e. in a maximum entropy state, thus inducing considerable initial mass-loss rates. In reality, a planet will cool and accrete gaseous material before disc dispersal, thus forming a radiative outer zone that limits cooling \citep[e.g.][]{Rafikov2006}. Furthermore, whilst an instantaneous disc dispersal is warranted given that $\tau_\text{disp} < \tau_\text{cool}$, the introduction of a finite dispersal timescale may allow the planet to remain in hydrostatic equilibrium with the disc for at least some of its evolution. This possibility is backed up by the fact that the mass-loss timescales found in \citet{Owen2016} were comparable to disc dispersal timescales.

This paper addresses many of the assumptions adopted in the aforementioned works. We use a self-consistent numerical evolution models model for gaseous accretion and boil-off that accounts for a non-instantaneous disc dispersal process. Beyond this, we then develop our models to include core luminosities and study the subsequent transition from boil-off to core-powered mass-loss \citep[e.g][]{Ginzburg2016,Gupta2019,Misener2021}. Finally, we consider stellar X-ray/EUV heating and the transition from boil-off or core-powered mass-loss to photoevaporation. We layout our numerical methods in Section \ref{sec:method}, with results in Section \ref{sec:results}, discussion in Section \ref{sec:discussion} and conclusions in Section \ref{sec:conclusion}.

\section{Method} \label{sec:method}

In order to model boil-off in a self-consistent manner, one must also consider the gas accretion phase that occurs before disc dispersal. Gas accretion is controlled by a planet's ability to cool \citep[e.g.][]{Lee2015}. Hence, the initial thermodynamic state of a planet at the onset of boil-off is affected by accretion. In \cite{Owen2016}, boil-off evolution models were performed using \textit{Modules for Experiments in Stellar Astrophysics}, \verb|MESA| \citep{Paxton2011,Paxton2013, Paxton2015, Paxton2018} and assumed planets undergo mass-loss via a hydrodynamic isothermal Parker wind \citep{Parker1958}. These models did not, however, self-consistently model the gas accretion phase before disc dispersal and instead assumed planets began from an initial inflated state in which $R_\text{ph} \sim R_\text{B}$. This can be thought of as a young, fully convective, maximum entropy planet that is inflated due to an accretion luminosity from the core's formation. 

As discussed, another assumption of \citet{Owen2016} was to assume an instantaneous disc dispersal process. Whilst the disc dispersal timescale is typically much shorter than the disc's lifetime, the disc will require a finite amount of time to reduce its gas surface density such that the planet is exposed to a vacuum.\footnote{Or strictly speaking, an external pressure sufficiently low to 
launch a transonic wind from the planet.} A gradual reduction in confining pressure will make the boil-off process less extreme since the planet has more time to respond thermodynamically. In this case, the planet may be able to remain in a quasi-hydrostatic equilibrium with the disc (implying causal contact), as was assumed in \citet{Ikomi2012}, for at least part of the planet's evolution.

The combination of these two assumptions: no gaseous accretion and hence maximally entropic initial planet, and instantaneous disc dispersal, imply that the mass-loss rates in \cite{Owen2016} were likely upper limits. The presence of an upper radiative zone and a non-instantaneous disc dispersal will act to reduce mass-loss.

\subsection{Atmospheric Structure Model} \label{sec:PlanetStructure}

Similar to \citet{Owen2016}, we perform evolution models with \verb|MESA| \citep{Paxton2011,Paxton2013, Paxton2015, Paxton2018}.  \verb|MESA| is an open-source one-dimensional stellar evolution code that can handle certain hydrodynamic processes. More recently, it has been used to understand planet evolution \citep[e.g.][]{Owen2013,Owen2016,ChenRogers2016,Kubyshkina2020,Kubyshkina2022,Malsky2020,Owen2020}. For a small mass exoplanet with H/He dominated atmosphere, it uses the SCVH equation of state from \cite{Saumon1995} and dust-free opacity tables from \cite{Freedman2008}. For each timestep in \verb|MESA|, the total mass in the simulation domain is adjusted following any mass-loss/accretion (as described in Section \ref{sec:MESA_massloss}). In addition to the timestep constraints within \verb|MESA|, we add the requirement that the change in the atmospheric mass fraction is $\leq 0.1\%$. The system then relaxes by re-solving the stellar structure equations, which include additional hydrodynamic terms to account for advection. 

An important limitation of \verb|MESA| is that it performs radiative transfer using the diffusion approximation, requiring the atmosphere to be locally optically thick in the entire domain. Specifically, this requires the photon mean-free path to be much shorter than the pressure scale height for the entire domain. In addition, the code can only solve the stellar structure equations in sub-sonic flow since the implicit numerical scheme can not accurately resolve the necessary sound waves to model shocks and supersonic flow. As discussed in Section \ref{sec:MESA_massloss}, we adopt a transonic outflow model that spans optically thin and thick regimes. In other words, the flow is launched in an optically thick region but extends beyond the photosphere. This limitation poses a severe challenge to modelling mass-loss. A solution exists, however, if one carefully considers where to place the outer boundary condition of the \verb|MESA| model.

\subsection{Outer Boundary Condition} \label{sec:MESA_BC}
\begin{figure} 
\begin{center}
\includegraphics[width=\columnwidth]{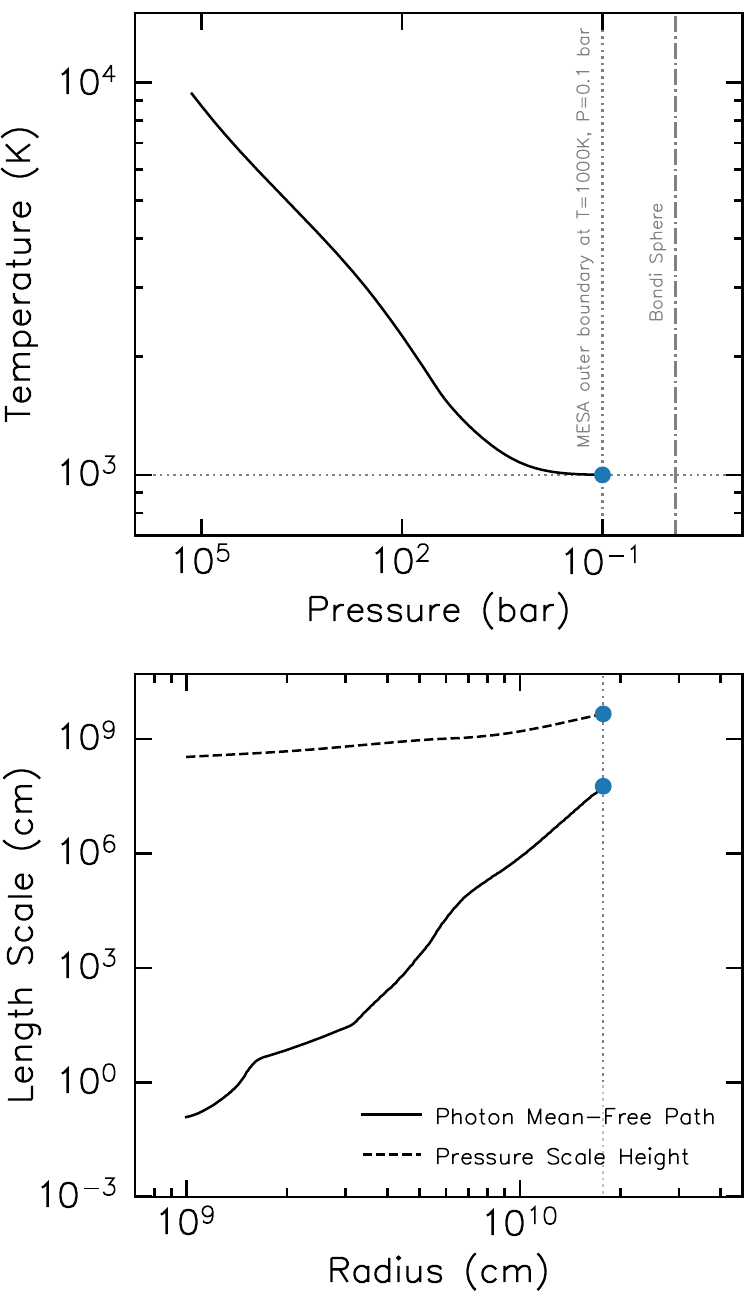}
    \caption{In the upper panel, a temperature-pressure profile is shown for an atmospheric model with $5M_\oplus$ core and atmospheric mass fraction of $\sim 10\%$ at $\sim 2$ Myr. The boundary condition (BC) is set with $P_\text{BC}=0.1 \text{ bar}$ and $T_\text{BC}=1000 \text{K}$, as shown with grey dashed lines and blue points. One can see that the temperature in the inner atmosphere is approximately adiabatic but flattens out in the outer radiative zone. Beyond this, the atmosphere is assumed to be isothermal. In addition, the location of the Bondi sphere is shown as a grey dot-dashed line. These properties imply that the MESA boundary condition is in an isothermal (radiative) and subsonic (interior to Bondi sphere) region. The lower panel shows the photon mean-free path, $\lambda_\text{ph} = 1/(\kappa \rho)$ (shown in black-solid), where $\kappa$ is gas opacity, which is much shorter than the pressure scale height (black-dashed) for all radii, meaning that the entire MESA domain is also locally optically thick.}
    \label{fig:BoundaryCond} 
\end{center}
\end{figure}

When modelling boil-off and hydrodynamic escape in general, one must consider supersonic flow. As discussed, \verb|MESA| sets two constraints determining where the simulation's outer boundary can be placed. Namely, the atmosphere below this point should be locally optically thick and subsonic. As described in the next Section, mass-loss is modelled via an isothermal Parker wind beyond this boundary, adding a third constraint: the atmosphere above the boundary should be approximately isothermal. To satisfy these constraints, we position the outer boundary condition at a fixed pressure $P_\text{BC}$ and temperature $T_\text{BC}$. The choice in pressure is arbitrary, although it is chosen to ensure the gas is locally optically thick and in the upper radiative (and therefore approximately isothermal) layer. For all models presented in this paper, we adopt a value of $P_\text{BC}=0.1$~bar, but confirm that this choice does not affect our results. To ensure that this pressure satisfies the required conditions, we show that the corresponding boundary condition radius $R_\text{BC}$ at $(P_\text{BC}, T_\text{BC})$ is beneath the Bondi radius (subsonic), in the radiative layer (isothermal) and that the photon mean-free path is much shorter than the pressure scale height for the entire domain (locally optically thick) in Figure \ref{fig:BoundaryCond}.

The outer boundary temperature, which is in the outer radiative layer of the atmosphere, is set to a constant equilibrium temperature $T_0=T_\text{eq}$ since it is in radiative equilibrium with the local protoplanetary disc and/or host star. In addition to a fixed temperature and pressure, \verb|MESA| also requires the optical depth to out-going radiation at this boundary. This relies on knowing the density structure above $R_\text{BC}$, which is determined by mass-loss/accretion, wich we discuss in the next Section. 

\subsection{Accretion and Mass-Loss} \label{sec:MESA_massloss}
In order to model gas accretion and boil-off for a planet immersed in a protoplanetary disc, we require the ability to model both accretion and mass-loss self-consistently. As discussed, mass-loss in \verb|MESA| is treated by adding/removing mass to the domain and then allowing the system to relax.\footnote{In practice, this mass-change is spread across a portion of the domain, predominantly in the upper atmosphere.} Given a set of boundary conditions (Section \ref{sec:MESA_BC}), one calculates and prescribes a mass-flux due to accretion/mass-loss. The question is, how to determine the mass-flux?

\begin{figure} 
\begin{center}
	\includegraphics[width=\columnwidth]{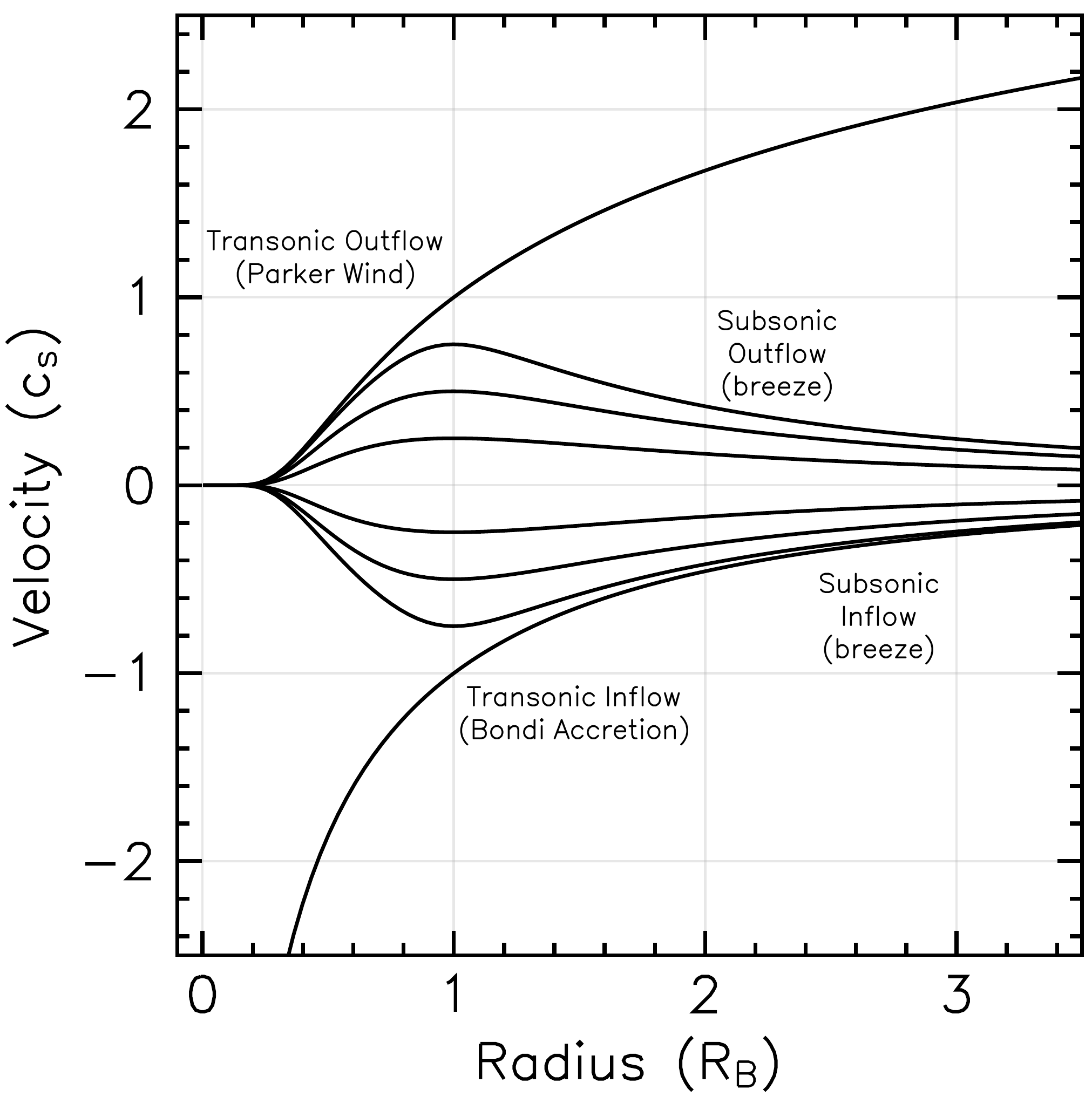}
    \caption{Velocity profiles for physical solutions to the isothermal fluid equations in non-hydrostatic equilibrium (see Equation \ref{eq:IsothermalODE}) are shown with radius. Here, six breeze solutions are shown, which remain subsonic for all locations ( $u/c_\text{s}<1$ for all $R/R_\text{B}$). Positive and negative velocities correspond to mass-loss and accretion, respectively. The two transonic solutions exceed the sound speed above/below the Bondi radius $R_\text{B}$. Planets can accrete/lose mass in our adopted evolution models via any of these solutions.}
    \label{fig:boilAccreteSolutions} 
\end{center}
\end{figure}

The distinction between mass-loss and accretion will depend on the physics of the disc. If the planet is pressure-supported by the local disc gas and can cool, then the atmosphere contracts, leading to an under-pressured system that cannot support the disc material above it, causing mass to flow onto the planet. On the other hand, if the disc's pressure support is insufficient, the upper atmosphere will expand and undergo mass-loss. We model the accreting/escaping atmosphere above the boundary condition with an isothermal layer, also set to $T_\text{eq}$, allowing one to solve the appropriate isothermal fluid equations to determine the flow structure and hence mass-flux into/out of the \verb|MESA| domain. The problem is set up as follows. As in the original Parker wind problem \citep{Parker1958}, one must solve the steady-state mass and momentum conservation equations:
\begin{equation} \label{eq:massCont}
    \frac{1}{r^2} \frac{\partial}{\partial r} (\rho u r^2) = 0  \implies \dot{M} = 4\pi\rho u r^2 = \text{const.}
\end{equation}
\begin{equation} \label{eq:steadystate-momentumcons}
    u \frac{\partial u}{\partial r} + \frac{1}{\rho} \frac{\partial P}{\partial r} + \frac{GM_\text{p}}{r^2} = 0,
\end{equation}
where $\dot{M}$ is the planet mass-loss rate. We adopt an ideal gas equation of state for constant temperature and combine to find:
\begin{equation}
    \frac{1}{\rho} \frac{\partial \rho}{\partial r} = -\frac{2}{r} - \frac{1}{u} \frac{\partial u}{\partial r},
\end{equation}
which is substituted into Equation \ref{eq:steadystate-momentumcons} to find the equation of motion for a steady state flow:
\begin{equation} \label{eq:IsothermalODE}
    \bigg( u - \frac{c_\text{s}^2}{u} \bigg) \frac{\partial u}{\partial r} = \frac{2 c_\text{s}^2}{r} - \frac{GM_\text{p}}{r^2}.
\end{equation}
This equation is trivially satisfied with $u = c_\text{s}$. In this case, the left-hand side becomes zero, allowing us to solve for the location of this critical point. One finds that the flow velocity is equal to the sound speed when:
\begin{equation}
    r = R_\text{B} \equiv \frac{GM_\text{p}}{2c_\text{s}^2},
\end{equation}
which represents the transonic Parker solution \citep{Parker1958}. Equation \ref{eq:IsothermalODE} is subject to an inner boundary condition provided by the outer boundary condition of the \verb|MESA| domain at $R_\text{BC}$, and an outer boundary condition which is controlled by the local gas density of the disc, $\rho_\text{disc}$. {\jr{Strictly speaking, the location of this boundary condition is $\text{min}(R_\text{B}, \, R_\text{Hill})$, where the Hill radius is defined as $R_\text{Hill} = a (M_\text{p} / 3M_*)^{1/3}$ for semi-major axis $a$ and stellar mass $M_*$, since these are the points where gas dynamics are controlled by the disc (and therefore the host star) and not the planet. We simplify the mathematics of the problem by always adopting the Bondi radius $R_\text{B}$ as our outer boundary location, which is justified for the vast majority of our parameter space range if one relates a planet's equilibrium temperature $T_\text{eq}$ to semi-major axis $a$ for a young, inflated pre-main sequence solar-mass star (as is done in Section \ref{sec:PhysicalDiscModels}).\footnote{{\jr{For example, from MIST stellar evolution models \citep{MIST-I2016}, a pre-main sequence solar-mass star at $1$~Myr has an effective temperature of $T_\text{eff} \sim 4500$~K and radius of $R_* \sim 2.5 R_\odot$, yielding a planetary equilibrium temperature of $T_\text{eq} = T_\text{eff} (R_* / 2a)^{1/2} \sim 1100$~K at $0.1$~AU. Combining with Equation \ref{eq:soundspeed}, these calculations yield a smaller Bondi radius than Hill radius for planets $\lesssim 6M_\oplus$ at $0.1$~AU and $\lesssim 9M_\oplus$ at $0.2$~AU.}}}}} As such, the density at the outer boundary for this isothermal layer is set to $\rho_\text{disc}$ at $R_\text{B}$. One can then relate $\rho_\text{disc}$ to the disc gas surface density $\Sigma_\text{g}$:
\begin{equation} \label{eq:rhodisc}
    \rho_\text{mid}(r) = \frac{\Sigma_\text{g}(r)}{\sqrt{2\pi} H} = \frac{\Sigma_\text{g}(r)}{c_\text{s}} \sqrt{\frac{GM_*}{2\pi a^3}},
\end{equation}
where $H = c_\text{s} / \Omega_\text{K}$ is the disc pressure scale height for a Keplerian angular frequency $\Omega_\text{K}$. As shown in \citet{Cranmer2004}, Equation \ref{eq:IsothermalODE} can be solved with Lambert W functions:
\begin{equation} \label{eq:u(r)}
    u(r) = c_\text{s} \sqrt{-W_k(-D(r))},
\end{equation}
where $k$ is the function-branch and $D(r)$ is:
\begin{equation} \label{eq:D(r)}
    D(r) = \exp \bigg \{ -4\ln r - 4\frac{R_\text{B}}{r} - \ln c_\text{s}^2 + C \bigg \}.
\end{equation}
Here, $C = 4 \ln R_\text{B} + 2 \ln u_\text{crit} - (u_\text{crit} / c_\text{s})^2 + 4$ is a constant and controlled by the fluid velocity, $u_\text{crit}$, at the critical point, $R_\text{B}$. If $|u_\text{crit}|<c_\text{s}$ then the solution is a subsonic breeze, whereas if $|u_\text{crit}|=c_\text{s}$, then the solution corresponds to in-flowing transonic Bondi accretion \citep{Bondi1952} or an out-flowing transonic wind \citep{Parker1958}. Examples of such solutions with varying critical velocity, $u_\text{crit}$, are shown in Figure \ref{fig:boilAccreteSolutions}. Breeze solutions remain subsonic for all radii, whilst transonic solutions exceed the sound speed beyond/interior to $R_\text{B}$. Note that for the transonic accretion solution, the velocity increases in magnitude as one approaches the planet and would eventually become supersonic and shock inside the \verb|MESA| domain. As discussed, \verb|MESA| cannot treat a supersonic fluid. However, since we are not modelling planets that enter a runaway accretion phase, the accretion occurs via breeze solutions that are sufficiently subsonic to not be of concern.

The difference between requiring a breeze solution or a transonic solution depends on the fluid pressure at the outer boundary \citep[e.g.][]{Velli1994,Keto2020}. In \citet{Owen2016}, only the transonic solution was considered due to the vacuum boundary condition and the inherent assumption of an instantaneous disc dispersal. In reality, the planetary flow will increase in launch velocity and transition from breeze solutions to the transonic wind as the disc gradually disperses. The faster this dispersal process happens, the quicker the outflow will tend to the transonic solution. In the case of hydrostatic equilibrium (HSE), the density profile in the isothermal layer beyond the \verb|MESA| domain is given by:
\begin{equation} \label{eq:HSEboilOff}
    \rho_\text{HSE}(r) = \rho_\text{BC} \exp \bigg \{ \frac{GM}{c_\text{s}^2} \bigg( \frac{1}{r} - \frac{1}{R_\text{BC}} \bigg) \bigg \}.
\end{equation}
Strictly speaking, this is a solution to the fluid equations if $u_\text{crit} = 0$. Now, if one manually decreases the density at the Bondi radius such that $\rho(R_\text{B}) < \rho_\text{HSE}(R_\text{B})$ (which in this problem is akin to reducing the gas density of the disc), then $u_\text{crit}$ will increase in an attempt to return to equilibrium. In the special case of an isothermal flow, the solution will remain as an out-flowing subsonic breeze if $e^{-0.5} \leq \rho(R_\text{B}) / \rho_\text{HSE}(R_\text{B}) < 1$ \citep{Lamers1999}. Suppose the density is reduced below this minimum value. In this case, the flow is no longer pressure-supported, the planet and disc lose their causal connection, and the only viable solution is a transonic wind. 

For this problem, it follows that for a given ratio of $\rho_\text{BC} / \rho_\text{disc}$ there exists a solution to the fluid equations in non-hydrostatic equilibrium with a unique $u_\text{crit} \in [-c_\text{s},+c_\text{s}]$. From mass continuity in Equation \ref{eq:massCont} at the two boundaries, we have:
\begin{equation}
    R_\text{BC}^2 \, \rho_\text{BC} \, u_\text{BC} = R_\text{B}^2 \, \rho_\text{disc} \, u_\text{crit}.
\end{equation}
Recall that there is only one unique solution to the isothermal fluid equations for a given $u_\text{crit}$. Hence one can calculate $u_\text{BC}$ for a given $u_\text{crit}$. Rearranging:
\begin{equation} \label{eq:outflowspeed}
    \frac{u_\text{BC}}{u_\text{crit}}  = \frac{\rho_\text{disc}}{\rho_\text{BC}} \, \bigg ( \frac{R_\text{B}}{R_\text{BC}} \bigg)^2,
\end{equation}
we may now solve this equation to find $u_\text{BC}$. Since $u_\text{BC}$ is a function of $u_\text{crit}$, this equation is solved numerically with a \verb|brentq| algorithm, adopting a numerical fractional tolerance of $10^{-12}$. Finally, the mass-flux $\dot{M}_\text{P}$ due to Parker-type solutions (including subsonic breezes) at the \verb|MESA| outer boundary is:
\begin{equation} \label{eq:breezeMassLoss}
    \dot{M}_\text{P} = 4 \pi R_\text{BC}^2 \rho_\text{BC} u_\text{BC}.
\end{equation}
Naively, one can adopt this value for each timestep of the \verb|MESA| model; however, this results in large, unstable oscillations in planet mass. This is because the value calculated for $\dot{M}_\text{P}$ in Equation \ref{eq:breezeMassLoss} is an \textit{instantaneous} mass-loss rate that is only appropriate for a set of boundary conditions at a given time. Recall, however, that \verb|MESA| is an implicit numerical scheme that takes large time steps. The value of $\dot{M}_\text{P}$ at the beginning of this timestep may, therefore, be very different to that at the end, particularly when noting that it is very sensitive to $\rho_\text{BC} / \rho_\text{disc}$. This sensitivity means that the value is a dramatic overestimate of the required value averaged over the timestep, meaning that the planet loses too much mass and then requires accretion to return to pressure support. This unstable oscillation continues, resulting in unphysical mass fluxes. We adapt \verb|MESA| with the following implicit mass-loss scheme to remedy this. 

Firstly, \verb|MESA| takes a step forward in time with $\dot{M} = 0$. The state of the system at the end of this timestep is then used to calculate $\dot{M}_\text{P}$, which represents the mass-loss that the Parker-type solutions can provide.\footnote{This value is in-fact averaged with that of the previous timestep to better estimate how $\dot{M}$ evolves with time.} Then, \verb|MESA| goes back and repeats the same timestep with iterative values of $\dot{M}$ until it converges on a value such that the system has returned to hydrostatic equilibrium with the disc i.e. $|(\rho(R_\text{B}) / \rho_\text{disc}) - 1| < \epsilon $, where $\epsilon = 10^{-3}$ is the numerical tolerance. This mass-loss rate is labelled as $\dot{M}_\text{HSE}$. Finally, \verb|MESA| performs the same timestep one more time, adopting the following mass-loss rate:
\begin{equation} \label{eq:implicitscheme}
    |\dot{M}| = \min \{ |\dot{M}_\text{P}|, |\dot{M}_\text{HSE}| \}.
\end{equation}
In other words, if the subsonic Parker-type solutions can provide enough mass-flux to return to hydrostatic equilibrium within each timestep, then \verb|MESA| adopts $\dot{M}_\text{HSE}$ and remains in causal contact with the disc. If this is not the case, then the system cannot return to equilibrium and $\dot{M}_\text{P}$ is adopted.

Now that the mass-loss rate is determined for each timestep, the density structure beyond the \verb|MESA| domain can be calculated, allowing one to evaluate the optical depth to out-going radiation as required in the \verb|MESA| outer boundary condition (see Section \ref{sec:MESA_BC}). For each evaluation of the model at a given timestep, the adopted value of $\dot{M}$ provides $u_\text{BC}$. Therefore, one can calculate the density profile in the isothermal layer from mass continuity:
\begin{equation}
    \rho(r) = \rho_\text{BC} \bigg( \frac{R_\text{BC}}{r} \bigg)^2  \frac{u_\text{BC}}{u(r)},
\end{equation}
where $u(r)$ is calculated using Lambert W function for a given $u_\text{BC}$. The optical depth contribution from this isothermal layer is thus:
\begin{equation} \label{eq:opticaldepth}
    \tau_\text{iso} = \int_{R_\text{BC}}^{R_\text{B}} \rho (r) \, \kappa(r) \, dr.
\end{equation}
Since \verb|MESA| utilises the dust-free opacities from \cite{Freedman2008} in its atmospheric structure calculations, we also adopt the same opacities for the optical depth calculation in Equation \ref{eq:opticaldepth} for consistency, evaluated at $T_\text{eq}$ and $\rho(r)$. This dust-free opacity choice is motivated by the fact that during inner disc dispersal, dust rapidly drains from the inner disc due to gas drag \citep[e.g.][]{Alexander2007,OwenKollmeier2019}. {\jr{We highlight that changing the gas opacities, such as the inclusion of dust, will act to change planetary cooling rates and thus the amount of atmospheric mass that is accreted onto the planet \citep[e.g.][]{Ikoma2000,Piso2015}. Regardless of this fact, however, boil-off will still operate since the process is triggered by a change in a planet's outer boundary condition, regardless of the exact amount of accreted gas.}} The contribution to optical depth from the disc is approximated to be $\tau_\text{disc} \approx \kappa_\text{disc} \Sigma_\text{g} / 2$ where $\kappa_\text{disc} = 0.01 \text{ cm}^2 \text{ g}^{-1}$ for a dust-free environment \citep[e.g.][]{D'Alessio2001}. Note that this choice in disc opacity has little effect on the evolution, although we comment on the potential effects of dust in Section \ref{sec:boiloffImprovements}. Finally, the total optical depth to out-going radiation beyond the \verb|MESA| boundary is:
\begin{equation}
    \tau_\text{BC} = \tau_\text{disc} + \tau_\text{iso} = \frac{\kappa_\text{disc} \Sigma_\text{g}}{2} +  \int_{R_\text{BC}}^{R_\text{B}} \rho (r) \, \kappa(r) \, \mathrm{d}r.
\end{equation}
In addition to contributing to optical depth, the isothermal layer contributes a small amount of atmospheric mass to the planet. Recall that the outer boundary $R_\text{BC}$ for the \verb|MESA| domain is defined at a fixed temperature and pressure, rendering it meaningless in terms of the physical mass of the planet. The true mass is calculated as the sum of the mass inside the \verb|MESA| domain and the mass in the isothermal layer out to the Bondi radius.

Note that our model differs from the work of \citet{Valletta2020} due to our outer boundary existing at fixed pressure and temperature. In the latter study, accretion was modelled for growing giant planets by considering an outer boundary at a fixed radius, with mass flux prescribed to ensure the latter condition is satisfied. As a result, there was no direct consideration of the fluid equations that govern accretion. Our approach allows for a self-consistent view of the interaction of a planet's atmosphere with its local protoplanetary disc, including the transition to mass-loss for low-mass planets.

\subsection{Initial MESA models}
Constructing initial models in \verb|MESA| for young planetary atmospheres is a convoluted process. Constructing a low-mass, low-luminosity and yet highly inflated body with a solid core requires multiple steps, which are outlined below and follow previous works of \citet[][]{Owen2013,ChenRogers2016,Kubyshkina2020,Kubyshkina2022,Malsky2020}.

\begin{enumerate}
    \item \textbf{Initial model}: To begin, the \verb|create_initial_model| routine is used to produce an initial H/He dominated model of $2M_\text{Jup}$ and $6R_\text{Jup}$. We choose a hydrogen mass fraction of 0.74, helium mass fraction of 0.24 and metal mass fraction of 0.02 to reflect typical nebular conditions. {\jr{This model is then evolved for $10^{12}$ yrs such that the planet has a low enough entropy in order for atmospheric mass to be removed stably in Step (iii). This allows one to produce a planet with the desired initial atmospheric mass fraction.}}
    
    \item \textbf{Add planetary core}: Next, the \verb|relax_core| routine is used to add an inert core of mass $M_\text{c}$. To determine the core radius $R_\text{c}$, we adopt the mass-radius relations of \citet{Fortney2007} for a composition of $1/3$ iron, $2/3$ silicate. Note that \verb|MESA| treats the core as inert and does not model its interior. In section \ref{sec:BO-CPML-XUV}, we introduce core luminosity due to cooling and radioactive decay. 
    
    \item \textbf{Choose initial atmospheric mass fraction}: To choose an atmospheric mass fraction $X$, the \verb|relax_mass| routine is used to reduce the initial $2M_\text{Jup}$ model to a desired value. Note, however, that this atmospheric mass fraction is arbitrary at this point since we allow the system to come into equilibrium with the disc and then accrete material prior to disc dispersal. The initial atmospheric mass fraction at the end of Step (v) is of physical interest to the problem.
    
    \item \textbf{Impose outer boundary condition}: At this point, we switch from the standard photospheric boundary condition implemented in \verb|MESA| \citep[see][]{Paxton2011} to one with fixed pressure and temperature as described in Section \ref{sec:MESA_BC}. This boundary condition is prescribed with the \verb|other_surface_PT| routine. At this point, we also perform tests to ensure this boundary is in an optically thick, subsonic and isothermal region (see Figure \ref{fig:BoundaryCond}).
    
    \item \textbf{Inflate and equilibrate}: To simulate a young planet in equilibrium with its nascent disc, one must increase the model entropy from the previous steps. We use the \verb|other_energy| routine to introduce a core luminosity that acts to inflate the planet, which can be interpreted as the accretion luminosity, $L_\text{acc}$, that would have been present during core formation. We iterate the core luminosity until the density at the Bondi radius is equal to that of the local disc density $|(\rho(R_\text{B}) / \rho_\text{disc}) - 1| < \epsilon $ where $\epsilon = 10^{-3}$ is the numerical tolerance.
    
    \item \textbf{Turn off core luminosity and accrete}: The final preparation step is to switch off the core luminosity from Step (v), reset the planet's age to a specified value $t_\text{pl}$ within the disc's lifetime, and allow gas accretion to commence, as described in Section \ref{sec:MESA_massloss}, which is implemented in the \verb|extras_check_model|. We additionally switch on the hydrodynamic fluid terms with \verb|change_v_flag| to allow for advection. The model can now self-consistently switch from accretion to mass-loss as the disc begins to disperse. At this point, which marks the onset of boil-off, we record the atmospheric mass fraction as $X_\text{init}$.
    
\end{enumerate}

The resulting model is representative of a young planet that has formed and accreted a H/He dominated atmosphere while immersed in a protoplanetary disc. The planet parameters at this point are the core mass, $M_\text{c}$, initial atmospheric mass fraction, $X_\text{init}$, cooling timescale $\tau_\text{cool}$, equilibrium temperature, $T_\text{eq}$, semi-major axis, $a$, and planet age $t_\text{pl}$. Note that this latter variable is, in fact, a crude estimate of the actual planet age, but instead represents the amount of time for which we have allowed accretion to take place. In essence, it is a variable that controls the initial atmospheric mass fraction of a planet before dispersal begins. 

\section{Results} \label{sec:results}
Our results are split into three parts: in Section \ref{sec:tcool_vs_tdisp} we aim to isolate the physics of boil-off and thus consider minimal disc models with planets at constant equilibrium temperature and no luminosity contribution from the core. In Section \ref{sec:BO-CPML-XUV} we introduce core luminosity and consider the transition from boil-off to core-powered mass-loss and/or XUV-driven photoevaporation. Finally, in Section \ref{sec:PhysicalDiscModels}, we employ more realistic disc models to investigate the effect of a changing equilibrium temperature throughout disc evolution. 

\begin{figure*}
\begin{center}
	\includegraphics[width=2.0\columnwidth]{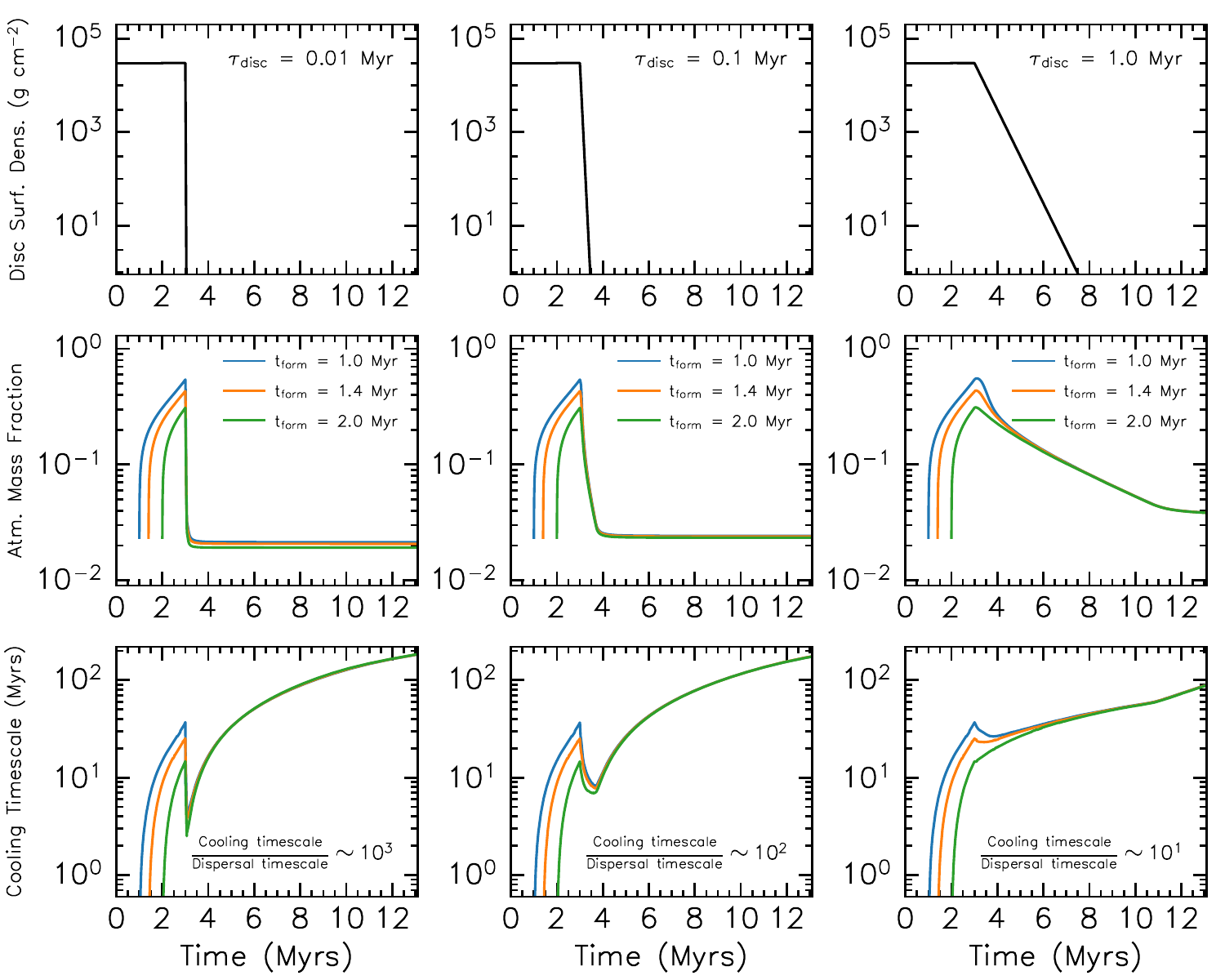}
    \caption{Accretion and boil-off are shown for planets with core masses of $5M_\oplus$ and constant equilibrium temperature of $900$K. The disc begins to disperse at $t_\text{disp}=3$~Myrs, with disc gas surface density shown in the top panels for initial surface density of $\Sigma_{0,\text{g}} = 3 \times 10^4 \text{\; g cm}^{-2}$ and dispersal timescales of $\tau_\text{disp} = 0.01$, $0.1$ and $1.0$~Myrs shown in left, middle and right-hand columns respectively. Atmospheric mass fraction is shown in the middle row, with planets forming at $1.0$, $1.4$ and $2.0$~Myrs shown in blue, orange and green respectively. The cooling timescale is shown in the bottom row, with its ratio to the disc dispersal timescale recorded at $3$~Myrs.}
    \label{fig:5Mearth900K_VariousDiscDisp} 
\end{center}
\end{figure*}

\begin{figure*} 
\begin{center}
	\includegraphics[width=2.0\columnwidth]{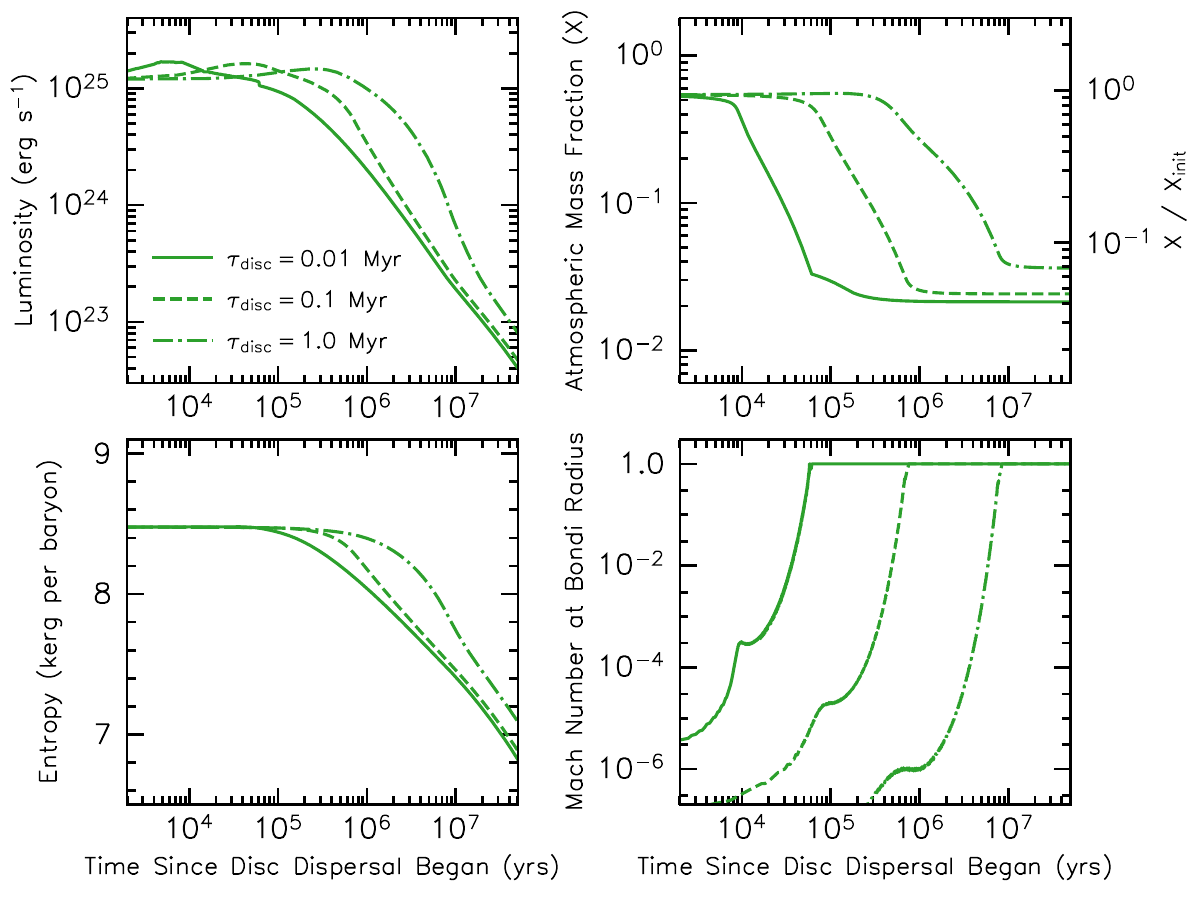}
    \caption{Boil-off is shown for planets with core masses of $5M_\oplus$ and constant equilibrium temperature of $900$K. Planets all begin evolution at $2$~Myrs, corresponding to green lines in Figure \ref{fig:5Mearth900K_VariousDiscDisp}. Time is measured from the point at which disc dispersal began, in this case at $t_\text{disp}=3 \times 10^6$ yrs. Upper-right shows the atmospheric mass fraction evolution, as in Figure \ref{fig:5Mearth900K_VariousDiscDisp}, additionally with the fractional change $X / X_\text{init}$. The lower-right panel shows the Mach number at the Bondi radius of the outflow, whilst the bottom-left shows the entropy at the base of the atmosphere. The upper-left panel shows the luminosity at the radiative-convective boundary. Solid, dashed and dot-dashed lines represent disc dispersal timescales of $\tau_\text{disp} = 10^4$, $10^5$ and $10^6$~yrs respectively.}
    \label{fig:5Mearth900K_Entropy} 
\end{center}
\end{figure*}

\subsection{Pure boil-off: the race between planet cooling timescale and disc dispersal timescale} \label{sec:tcool_vs_tdisp}

To begin, we evolve the local disc gas surface density $\Sigma_\text{g}$ with the following parameterisation:
\begin{equation} \label{eq:discdispersal}
    \Sigma_\text{g}= \begin{cases}
    \Sigma_{\text{g},0} & t < t_\text{disp} \\
    \Sigma_{\text{g},0} \, \exp \{ \frac{1}{\tau_\text{disp}} (t_\text{disp} - t) \} & t \geq t_\text{disp},
    \end{cases}
\end{equation}
where $\Sigma_{\text{g},0}$ is the initial gas surface density, $t_\text{disp}$ is the time at which the disc starts to disperse and $\tau_\text{disp}$ is the dispersal timescale. To investigate the effect of disc dispersal rate, Figure \ref{fig:5Mearth900K_VariousDiscDisp} shows the evolution of a $5M_\oplus$ core at $0.1$AU, with a constant equilibrium temperature of $900$K. The disc starts dispersing at $t_\text{disp}=3 \times 10^6$ yrs, but with three separate dispersal timescales: $\tau_\text{disp} = 10^4\text{, } 10^5\text{ and } 10^6 \text{ yrs}$. In addition, we begin planet evolution at different times: $1 \times 10^6$, $1.4 \times 10^6$ and $2 \times 10^6\text{ yrs}$. As a result, these planets have a range in atmospheric mass fractions at the point of disc dispersal since they have had varying amounts of time to cool and accrete. Figure \ref{fig:5Mearth900K_VariousDiscDisp} reveals that, despite having initial atmospheric mass fractions between $30-50\%$ at the onset of disc dispersal, the final atmospheric mass fraction is significantly reduced by greater than $\sim 90\%$. Planets that undergo dispersal timescales of $\tau_\text{disp} = 10^4$, $10^5$ and $10^6$~yrs are left with $\sim 2\%$, $\sim 2.3\%$ and $\sim 4\%$ atmospheric mass fractions respectively. Thus, for a given core mass, equilibrium temperature and dispersal timescale, planets all host approximately the same atmospheric mass fraction, independently of how much atmosphere they initially accreted. The key quantities controlling mass-loss are the disc dispersal timescale $\tau_\text{disp}$, and the atmosphere's cooling timescale  $\tau_\text{cool}$ (which is a proxy for planet entropy). This latter variable is important since it controls the phase of accretion that occurred prior to boil-off and ultimately sets the ability of the planet to thermally respond to the changing disc boundary condition. In the case of no internal energy sources (such as a cooling planetary core, as considered in Section \ref{sec:BO-CPML}), the cooling timescale is the time it would take to radiate the total energy $E_\text{tot}$ of the atmosphere away, with contributions from internal thermal energy and gravitational binding energy:
\begin{equation} \label{eq:Cooltimescale}
    \tau_\text{cool} = \frac{|E_\text{tot}|}{L} = \frac{1}{L} \; \bigg| \int_{M_\text{c}}^{M_\text{p}} e_\text{i}(r) - \frac{G m(<r)}{r} \; \mathrm{d}m \; \bigg|,
\end{equation}
where $e_\text{i}$ is the internal thermal energy, $m(<r)$ is the mass interior to radius $r$, and $L$ is the radiative luminosity of the planet, evaluated at the radiative-convective boundary. If the ratio of $\tau_\text{cool} / \tau_\text{disp}$ is large, the planet cannot thermally respond to the disc dispersing, promoting adiabatic expansion and increased mass-loss rates. In the bottom panel of Figure \ref{fig:5Mearth900K_VariousDiscDisp} we also show the cooling timescale of the planet, which displays a sudden drop at the point of boil-off due to the sharp decrease in atmospheric mass. In Figure \ref{fig:5Mearth900K_VariousDiscDisp} we also note the ratio of cooling timescale to dispersal timescale, recorded at the time of disc dispersal onset at $3\times 10^6$~yrs. The larger this ratio, the more mass is lost. We note that in the case of very rapid disc dispersal $\tau_\text{disp} = 10^4 \text{ yrs}$, planets that accreted more atmosphere retain a slightly more massive atmosphere by the end of boil-off. 

A different regime exists, however, if the ratio of cooling timescale to dispersal timescale becomes less than unity. In this case, the planet can remain in thermodynamic equilibrium with the disc, and would not undergo the significant mass-loss and associated cooling processes involved in boil-off. Whilst a small amount of accretion would continue after the disc begins to disperse, the planet would then release this atmosphere to stay in equilibrium with a vanishing disc. This is indeed the case considered in \citet{Ikomi2012}, which resulted in very small atmospheres after dispersal concluded. One can see that in the case of the longest dispersal timescale in Figure \ref{fig:5Mearth900K_VariousDiscDisp}, a small amount of accretion occurs after the onset of disc dispersal at $3$~Myrs.

The results in Figure \ref{fig:5Mearth900K_VariousDiscDisp} suggest that in the case of a long cooling timescale compared to dispersal timescale (as is suggested by the disc observations), boil-off sets the initial conditions for the post-disc thermal evolution of close-in exoplanets. To understand this further, Figure \ref{fig:5Mearth900K_Entropy} shows the evolution of Mach number at the Bondi radius and entropy at the base of the atmosphere for the case of planets beginning evolution at $2 \times 10^6\text{ yrs}$ (green lines) from Figure \ref{fig:5Mearth900K_VariousDiscDisp}. The Mach number evolution demonstrates that the disc dispersal timescale controls the boil-off process. Initially, boil-off presents as a low Mach number outflow while the planet and disc can maintain causal contact via an instantaneous pressure-support mediated through mass-loss (see Equation \ref{eq:implicitscheme}). However, as the disc density drops further, the Mach number must increase to maintain pressure support at a rate that is controlled by the dispersal timescale. Eventually, the Mach number must rise so high that a causal connection cannot be maintained and a transonic wind is launched with unity Mach number.

Intuitively, the disc dispersal rate must control the outflow speed since the ratio of planet density and disc density controls the Mach number and therefore mass-loss rate (see Equation \ref{eq:outflowspeed}). If a planet were to have too much mass removed at a given time, then mass-loss would slow until the disc disperses to a sufficient level. Mass-loss would then resume at the rate determined by disc dispersal. Therefore, this highlights a significant result from this work: breeze solutions allow a planet to remain in causal contact with the local disc, allowing the disc dispersal process to dictate the final atmospheric mass fraction. In essence, this then builds upon the separate effects from \citet{Ikomi2012} and \citet{Owen2016}, which we discuss in Section \ref{sec:breezes}.

Figure \ref{fig:5Mearth900K_Entropy} also shows the entropy at the base of the atmosphere as a function of time after disc dispersal begins. Planets that undergo a more rapid disc dispersal process end with a lower final entropy, implying they have cooled more dramatically. To understand this, we also show the luminosity at the radiative-convective boundary (RCB) $L_\text{rcb}$ as a function of time. Planets experiencing a quicker disc dispersal show an increase in $L_\text{rcb}$ at the onset of dispersal. This effect is due to a brief period of $P \textrm{d} V$ expansion, which sets up a steep temperature gradient, thus enhancing radiative cooling. As previously highlighted, the magnitude of the $P \textrm{d} V$ expansion and, thus, radiative cooling is based on the balance of cooling timescales and dispersal timescales. For planets with a larger ratio of $\tau_\text{cool} / \tau_\text{disp}$, the greater the magnitude of $P \textrm{d} V$ expansion since the initial evolution is better approximated as adiabatic. Nevertheless, and as initially found in \citet{Owen2016}, we confirm that advective cooling is the dominant source of energy transport deep in the atmosphere for rapid disc dispersal. 

\subsection{The interplay between boil-off, core-powered mass-loss and XUV photoevaporation} \label{sec:BO-CPML-XUV}
Whereas in Section \ref{sec:tcool_vs_tdisp}, we did not consider any luminosity emanating from the planetary core; in reality, the base of the atmosphere will interact with a young molten surface and cool as a coupled system. To address this, we follow \citet{Lopez2012,ChenRogers2016,Vazan2018} and assume the core is isothermal with temperature $T_c$, which is initially set equal to the base temperature of the atmosphere (which is determined through \verb|MESA|'s structure model). We quantify the core cooling luminosity $L_\text{cool}$ as:
\begin{equation}
    L_\text{cool} = - c_\text{v} M_\text{core} \frac{\mathrm{d} T_c}{\mathrm{d} t},
\end{equation}
where we assume a heat capacity appropriate for a silicate mantle $c_\text{V}=1.0 \text{ J K}^{-1} \text{ g}^{-1}$ \citep[e.g.][]{Alfe2002,Valencia2010}. Numerically determining $\mathrm{d} T_\text{c} / \mathrm{d}t$ is not straightforward during rapid changes in an atmospheric structure such as those involved in boil-off. The core acts as a significant reservoir of thermal energy due to its large heat capacity, meaning that only a small change in core temperature $T_\text{c}$ over the timescales that boil-off acts can deposit significant energy into the atmosphere, which poses a complex numerical challenge (particularly for an implicit integration scheme). To address this, we exploit the notion that the magma surface is very unlikely to remain in thermal equilibrium with the base of the atmosphere on the very short timescales that boil-off operates $\lesssim 10^5$~yrs. In fact, \citet{Markham2022,Misener2022,Misener2023} have shown that atmospheric convection may be inhibited near the core due to high mean-molecular weight gradients, leading to an interior radiative zone. One can speculate that the inefficient cooling of such a radiative zone may provide a mechanism for a planet's core and atmosphere to cool out of equilibrium. Nevertheless, in the absence of constraints on the equilibration timescales of magma-atmosphere boundaries, we parameterise $\mathrm{d} T_\text{c} / \mathrm{d}t$ with a Newtonian cooling term:
\begin{equation} \label{eq:NewtonianCoreCooling}
     \frac{\mathrm{d} T_c}{\mathrm{d} t} = \frac{T_\text{atm,base} - T_\text{c}}{\tau_\text{c}},
\end{equation}
where $T_\text{atm,base}$ is the temperature at the base of the atmosphere and $\tau_\text{c}$ is the thermal equilibration timescale of the core-atmosphere boundary. During periods of dramatic atmospheric cooling due to mass-loss, we adopt core temperature evolution according to Equation \ref{eq:NewtonianCoreCooling}. Once the core and atmospheric base have returned to equilibrium, we set $T_\text{c} = T_\text{atm,base}$ such that $\mathrm{d} T_\text{c} / \mathrm{d}t = \mathrm{d} T_\text{atm,base} / \mathrm{d}t$ for the remaining evolution. We find that unphysical numerical solutions exist for $\tau_\text{c} \lesssim 10^4$~yrs, although we do not imply that physical solutions do not exist below this value, merely that our numerical scheme cannot find them. In the range $10^4 \lesssim \tau_\text{c} \lesssim 10^7$~yrs, we find that atmospheric evolution is weakly sensitive to $\tau_\text{c}$. We therefore set the equilibration timescale to $\tau_\text{c} = 10^5$~yrs for all models. We note, however, that if $\tau_\text{c} \gtrsim 10^7$~yrs, then the core-atmosphere boundary remains out of equilibrium until after boil-off has concluded, leading to a different evolution. In this case, we reproduce the results of \citet{Vazan2018} with delayed planet contraction over timescales of $\tau_\text{c}$.

In addition to core cooling, heating also acts as a result of radioactive decay in the core $L_\text{radio}$, for which we consider the contributions of $^{238}$U, $^{235}$U, $^{232}$Th, $^{40}$K and $^{26}$Al, which have half-lives of $4.5$ Gyrs, $0.7$ Gyrs, $14.1$ Gyrs, $1.3$ Gyrs and $0.7$ Myrs respectively. We adopt meteoritic abundances for all nuclides from \citet{Anders1989}, apart from $^{26}$Al, which is taken from \citet{Lee1976,Lee1977}. This latter isotope dominates radiogenic heating during boil-off due to its short half-life. Nevertheless, the cooling luminosity is typically orders of magnitude larger than that of radioactivity. The total core luminosity is given by:
\begin{equation}
    L_\text{core} = L_\text{cool} + L_\text{radio}.
\end{equation}
The effect of core heating is to increase mass-loss rates since the additional luminosity supplies more material to the base of the outflow (or, in other words, increases the atmosphere density at the base of the outflow).

\subsubsection{Transition from boil-off to core-powered mass-loss} \label{sec:BO-CPML}

\begin{figure}
\begin{center}
	\includegraphics[width=\columnwidth]{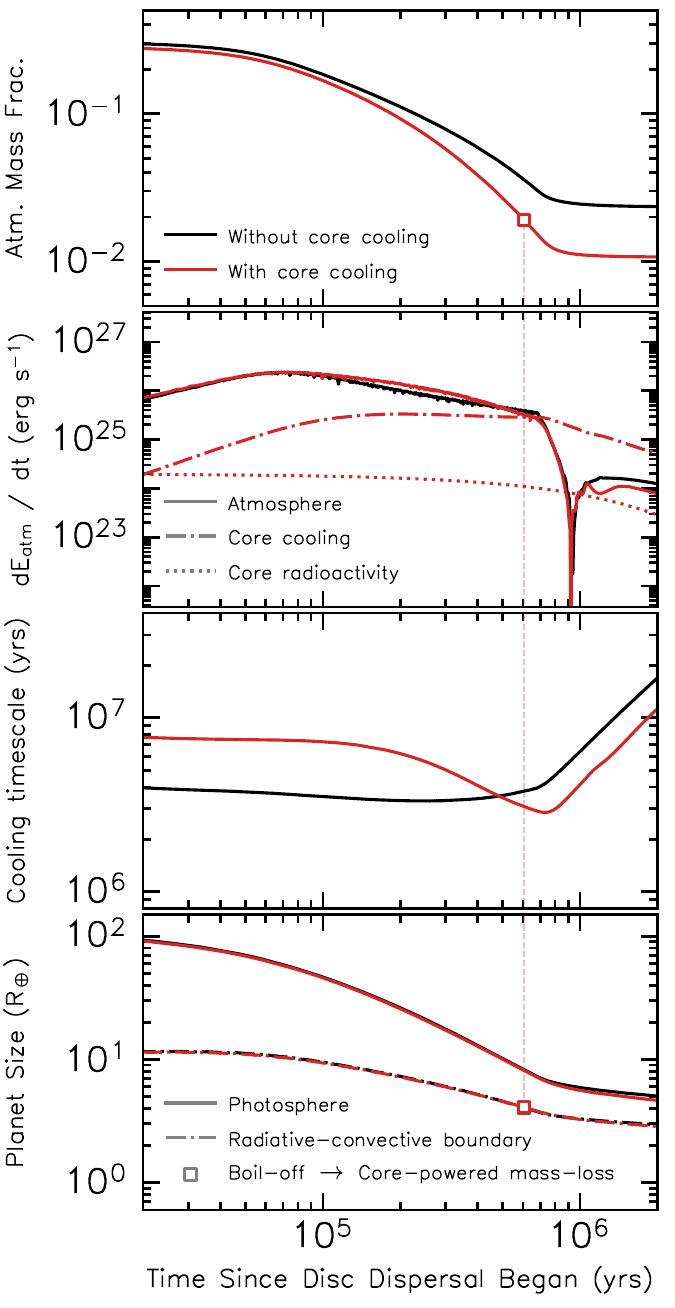}
    \caption{The effect of core luminosity for a $5M_\oplus$ core with constant equilibrium temperature of $900$K and disc dispersal timescale of $\tau_\text{disp}=10^5$~yrs. Time is measured from the point at which disc dispersal began, at $t_\text{disp}=3 \times 10^6$ yrs. Atmospheric mass fraction evolution is shown in the top panel with and without core cooling in black and red, respectively. The second panel shows the rate of change in atmospheric energy $\dot{E}_\text{atm} \equiv \mathrm{d} E_\text{atm} / \mathrm{d}t$ (as defined in Equation \ref{eq:Latm}) with solid lines. Note that $\dot{E}_\text{atm}$ is initially positive and then changes sign to be negative $\sim 9 \times 10^5$~yrs after disc dispersal began. The core luminosity, including contributions from core cooling $L_\text{cool}$ and radiogenic heating $L_\text{radio}$, is shown in dot-dashed and dotted, respectively. In the third panel, the cooling timescale is shown, as defined in Equations \ref{eq:Cooltimescale} and \ref{eq:coolingtimescale_wCore} for the case of with and without core luminosity, respectively. Finally, planet radius evolution is shown in the bottom panel, with the photospheric radius and radiative-convective boundary shown in solid and dot-dashed, respectively. The transition between boil-off and core-powered mass-loss is defined as the point at which the core luminosity becomes greater than the rate of change of atmospheric energy $L_\text{core} \geq \dot{E}_\text{atm}$. This point is shown for each model with coloured squares.}
    \label{fig:LcoreEffects} 
\end{center}
\end{figure}

The introduction of core luminosity allows us to consider the process of core-powered mass-loss in detail \citep{Ginzburg2018,Gupta2019,Gupta2020,Gupta2021,Gupta2022,Misener2021}. It is crucial to note that the physics of boil-off and core-powered mass-loss are fundamentally linked. Both processes are driven by bolometric heating of a planet's upper atmosphere, with mass-loss throttled by the rate at which material can be removed from the planet via a thermally driven Parker wind/breeze. However, energy is required to transport material from deep in the planet's gravitational well to the base of the outflow. During boil-off, transport energy is supplied via advection, as well as rapid cooling and contraction of the planet's atmosphere, which thus releases binding energy. On the other hand, core-powered mass-loss relies on core luminosity to transport material to the base of the outflow. In other words, the removal of mass is not `core-powered'; instead, the core luminosity allows material to be supplied to the top of the atmosphere, wherein it is heated and physically removed by the host star. This effect is precisely why core-powered mass-loss predicts a radius gap with dependence on incident stellar flux \citep{Gupta2019}. 

Figure \ref{fig:LcoreEffects} demonstrates the effects of adding core luminosity. We run evolution models with a $5M_\oplus$ core with constant equilibrium temperature of $900$K, disc dispersal onset at $t_\text{disp}=3 \times 10^6$~yrs and dispersal timescale of $\tau_\text{disp}=10^5$~yrs. We compare the scenario with and without core luminosity in black and red, respectively. In the top panel, we show the evolution of atmospheric mass fraction, highlighting that allowing the core and atmosphere to cool as a coupled system enhances mass-loss. 

In the second panel, we show the rate of change in atmospheric energy as solid lines, with core luminosities from cooling and radiogenic heating in dot-dashed and dotted, respectively. We demonstrate that core cooling dominates over radiogenic heating, as found in similar works \citep[e.g.][]{Lopez2012,ChenRogers2016}. Note that the definition of the rate of change in total atmospheric energy is as follows:
\begin{equation} \label{eq:Latm}
    \dot{E}_\text{atm} \equiv \frac{\mathrm{d} E_\text{atm}}{\mathrm{d} t} = \frac{\mathrm{d}}{\mathrm{d}t} \; \int_{M_\text{c}}^{M_\text{pl}} e_\text{i}(r) - \frac{G m(<r)}{r} \; \mathrm{d} m .
\end{equation}
In other words, $\dot{E}_\text{atm}$ is the rate of change in total atmospheric energy, which includes change in internal energy, where $e_\text{i}$ is the internal specific energy; change in binding energy, where $m(<r)$ is the mass interior to radius $r$; and change in mass, where $M_\text{pl}$ is the total mass of the planet (which, crucially, changes with time) interior to the Bondi radius. This definition differs from the planet luminosity shown in Figure \ref{fig:5Mearth900K_Entropy}, representing the radiative luminosity released at the radiative-convective boundary. In the case of no mass-loss, the two definitions are identical since energy is only lost due to radiation. During mass-loss, the planet loses negative gravitational potential energy, hence $\dot{E}_\text{atm} > 0$. Once mass-loss ceases, the planet loses energy due to cooling and gravitational contraction. As a result, the $\dot{E}_\text{atm} < 0$ after $\sim 9 \times 10^5$~yrs, which can be seen as a change in sign in Figure \ref{fig:LcoreEffects}.

With the planetary energy budgets in mind, we choose to define the transition between boil-off and core-powered mass-loss as the point at which an atmosphere's rate of change in energy (as defined in Equation \ref{eq:Latm}) becomes less than the luminosity released from the core. In order words, core-powered mass-loss begins when the core dominates the planetary energy budget. Mathematically, this transition occurs when $L_\text{core} \geq \dot{E}_\text{atm}$, which we highlight in Figure \ref{fig:LcoreEffects} with a square marker. Note, however, that the bottleneck for both boil-off and core-powered mass-loss is the same, being the Parker-wind mass-loss rates from Equation \ref{eq:breezeMassLoss}. Under some circumstances, the energy liberated from the core can provide the limit for mass-loss \citep[e.g.][]{Ginzburg2018}; however, we do not find any examples of such cases. As a result, we do not expect the behaviour of atmospheric escape to change dramatically during the transition from boil-off to core-powered mass-loss. 

The third panel of Figure \ref{fig:LcoreEffects} shows the cooling timescale for the planet's atmosphere with and without the influence of core luminosity. In the case of planetary evolution with core luminosity, the definition of cooling timescale changes from that of Equation \ref{eq:Cooltimescale}, to:
\begin{equation} \label{eq:coolingtimescale_wCore}
\begin{split}
    \tau_\text{cool} & = \frac{E_\text{c} + |E_\text{atm}|}{L} \\
    & = \frac{1}{L} \bigg\{ c_\text{v} M_\text{c} T_\text{c} + \bigg| \int_{M_\text{c}}^{M_\text{p}} e_\text{i}(r) - \frac{G m(<r)}{r} \; \mathrm{d}m \; \bigg| \;\bigg\},
\end{split}
\end{equation}
where we have additionally accounted for the thermal energy reservoir stored in the planetary core $E_\text{c}$. Note that here we approximate the core as incompressible (as is assumed in our \verb|MESA| model), meaning that the change in binding energy of the core does not contribute significantly to the thermal evolution of the planet. As with Equation \ref{eq:Cooltimescale}, $E_\text{atm}$ is the atmosphere's total energy and $L$ is the radiative luminosity at the radiative-convective boundary. Initially, the cooling timescale is larger in the case of a cooling core (red line) since the planet has a larger energy reservoir. As mass-loss continues, however, the planet hosts a less massive atmosphere and has a higher radiative luminosity due to the contributions from the core, meaning that the cooling timescale drops to lower values during boil-off than in the case of no core-cooling. Once mass-loss has concluded, we still find that planets have prematurely cooled and display a longer cooling timescale than their age, as identified in \citet{Owen2016}. Finally, the bottom panel of Figure \ref{fig:LcoreEffects} shows the planet's photospheric radius and radiative-convective boundary evolution in solid and dot-dashed lines, respectively. Note, however, that prior to the disc completely dispersing (coinciding with $\sim 10^6$~yrs after disc dispersal began in Figure \ref{fig:LcoreEffects}), the photospheric radius of a planet is physically meaningless.

\subsubsection{Transition from boil-off or core-powered mass-loss to XUV photoevaporation} \label{sec:BO/CPML-PE}

\begin{figure*} 
\begin{center}
	\includegraphics[width=2.0\columnwidth]{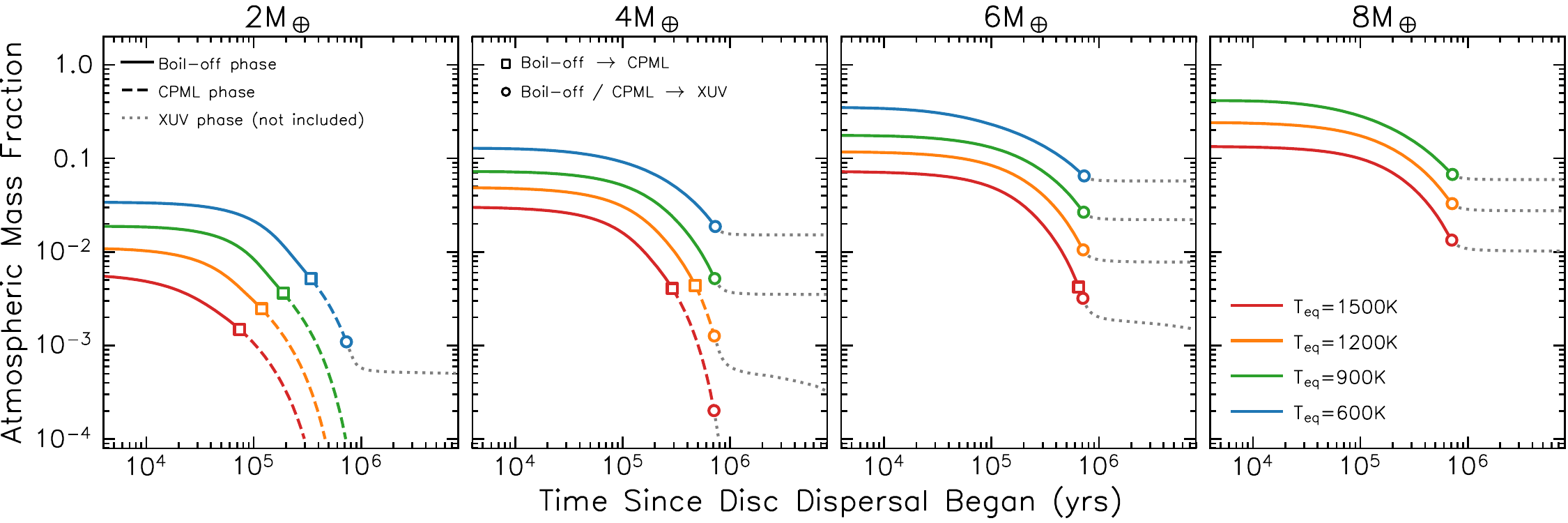}
    \caption{Atmospheric mass fraction evolution is shown for planets with core masses of $2M_\oplus$, $4M_\oplus$, $6M_\oplus$ and $8M_\oplus$ (moving from left to right) at constant equilibrium temperatures of $600$K, $900$K, $1200$K, $1500$K in blue, green, orange and red respectively. Time is measured from the point at which disc dispersal began, in this case at $t_\text{disp}=3 \times 10^6$ yrs. We assume a disc dispersal timescale of $\tau_\text{disp} = 10^5$~yrs. Lines demonstrate the transition from boil-off driven mass-loss (solid lines), to core-powered mass-loss (CPML; dashed lines, with transition denoted with squares) and XUV-driven photoevaporation (grey dotted, with transition denoted with circles). Note that although we mark the transition to photoevaporative mass-loss as circles in the models, we do not include the relevant photoevaporative mass-loss rates after this point in our models.}
    \label{fig:mass_grid_constT} 
\end{center}
\end{figure*}

Whereas boil-off and core-powered mass-loss rely on bolometric heating of escaping atmosphere, photoevaporation relies on X-ray/EUV (XUV) photons to drive hydrodynamic outflow \citep[e.g.][]{Erkaev2007,MurrayClay2009,OwenJackson2012}. The transition between these heating regimes is characterised by the depth of XUV photon penetration into the planet's atmosphere. As shown in \citet{Owen2023b}, photoevaporation dominates over a bolometric-heated wind (such as boil-off or core-powered mass-loss) when XUV photons penetrate below its sonic surface of the bolometrically heated region at $R_\text{B}$. This allows the gas to be heated to typical temperatures of $\sim 10^4$~K in the subsonic (and thus causally connected) portion of the outflow. To quantify this effect, we follow \citet{Owen2023b} in determining the XUV penetration radius $R_\text{XUV}$, which occurs at typical column (number) densities of $N \approx 10^{22}$~cm$^{-2}$ \citep[e.g.][]{Ercolano2009} for soft X-rays, which dominate the mass-loss at early ages \citep[e.g.][]{OwenJackson2012,Owen2013}. Hence, we state that the following is true at $R_\text{XUV}$:
\begin{equation} \label{eq:XUV1}
    N_\text{XUV} = \int_a^{R_\text{XUV}} n(r) \; \mathrm{d}r = 10^{22} \; \text{cm}^{-2},
\end{equation}
where $n(r)$ is the gas number density between the planet and star, including contributions from the atmosphere and protoplanetary disc. Beyond $R_\text{XUV}$, we assume the outflow is isothermal with a temperature of $T_\text{XUV} = 10^4$~K and accelerated to the sound speed $c_\text{s,XUV} = (k_\text{B} T_\text{XUV} / \mu m_\text{H})^{1/2}$. In order to determine the density at the boundary between bolometric and XUV-heated regions, we apply conservation of momentum at $R_\text{XUV}$, yielding:
\begin{equation} \label{eq:XUV2}
    \rho_\text{XUV}(R_\text{XUV}) = \rho_\text{bol}(R_\text{XUV}) \frac{c_\text{s}^2}{2c_\text{s,\text{XUV}}^2},
\end{equation}
where subscripts `XUV' and `bol' refer to variables in the XUV and bolometric heated region, respectively. Here, we have neglected the ram pressure term in the bolometric heated region since the flow is sufficiently subsonic ($u_\text{bol} / c_\text{s} \ll 1$). We also assume the flow in the XUV heated region is transonic at $R_\text{XUV}$, hence $u_\text{XUV}(R_\text{XUV}) \approx c_\text{s,\text{XUV}}$. The density profile in the XUV heated region beyond $R_\text{XUV}$ is calculated with:
\begin{equation} \label{eq:XUV3}
    \rho(r > R_\text{XUV}) = \rho_\text{XUV}(R_\text{XUV}) \bigg( \frac{R_\text{XUV}}{r} \bigg)^2  \frac{c_\text{s,\text{XUV}}}{u_\text{XUV}(r)},
\end{equation}
% \begin{equation} \label{eq:XUV3}
%     \rho(r > R_\text{XUV}) = \rho_\text{XUV}(R_\text{XUV}) \frac{R_\text{XUV}^2}{r^2}.
% \end{equation}
where, in order to calculate $u_\text{XUV}(r)$, we set $u = c_\text{s,\text{XUV}}$ at $R_\text{XUV}$ in the general solutions to isothermal outflow from Equations \ref{eq:u(r)} and \ref{eq:D(r)}. Equations \ref{eq:XUV1}, \ref{eq:XUV2} and \ref{eq:XUV3} are solved simultaneously with a \verb|brentq| solver at each timestep of our evolution calculations to find $R_\text{XUV}$, adopting a numerical fractional tolerance of $10^{-12}$. As in \citet{Owen2023b}, we assert that the planetary outflow is dominated by photoevaporation over boil-off or core-powered mass-loss if $R_\text{XUV} < R_\text{B}$. Note that explicitly including the photoevaporative-driven mass-loss rates beyond this transition is beyond the scope of this work and are therefore not included in our evolutionary calculations. Instead, we record planet properties at the point of XUV penetration and consider these as the initial conditions for photoevaporation models.

\begin{figure*} 
\begin{center}
	\includegraphics[width=2\columnwidth]{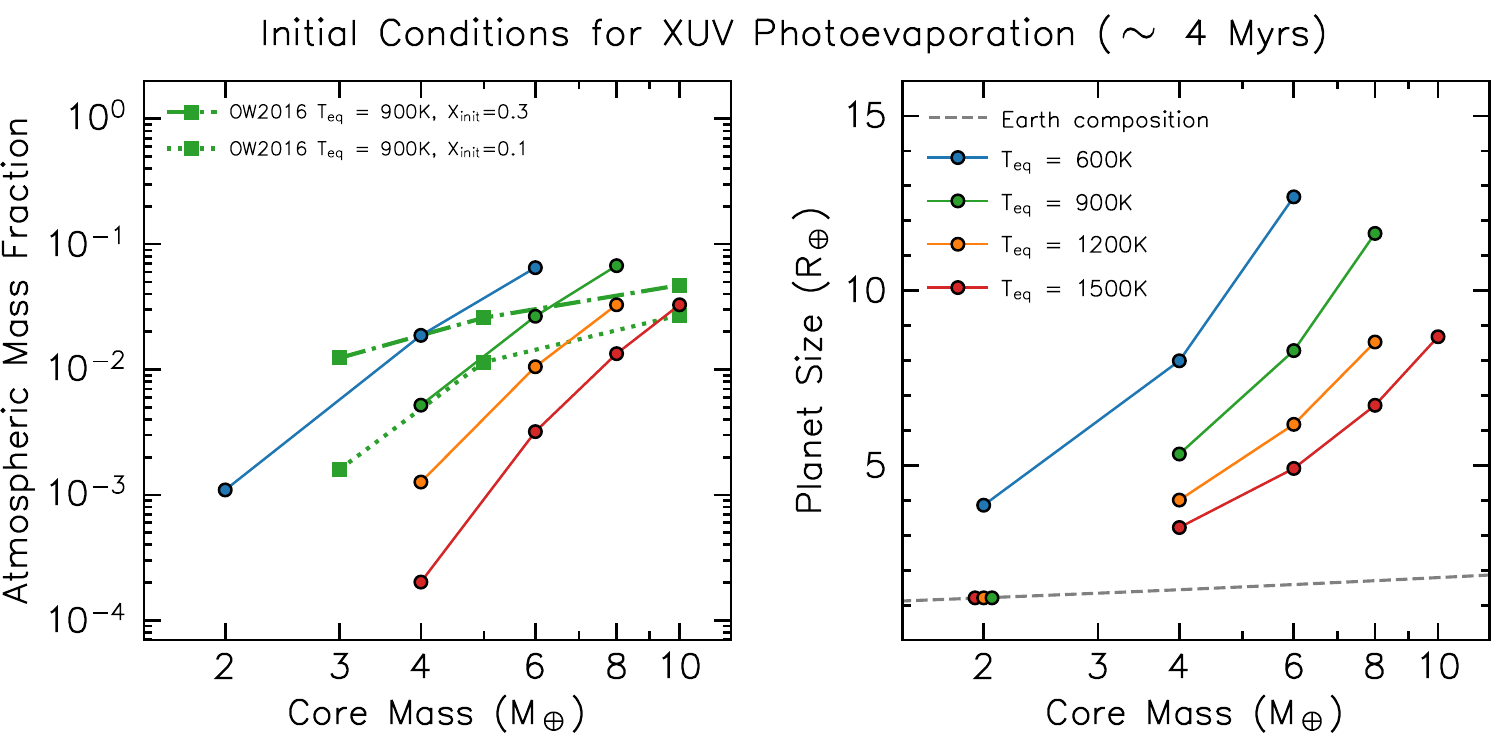}
    \caption{A summary of planet properties at the onset of XUV photoevaporation from models in Figure \ref{fig:mass_grid_constT} with constant equilibrium temperatures of $600$K, $900$K, $1200$K, and $1500$K in blue, green, orange and red respectively. We assume the disc disperses at $t_\text{disp}=3 \times 10^6$ yrs with a disc dispersal timescale of $\tau_\text{disp} = 10^5$~yrs, all of which lead to XUV photoevaporation dominating at $\sim 4$~Myrs, which coincides with the local protoplanetary disc having completely dispersed in these models. The left-hand panel shows the atmospheric mass fraction. Green lines with squares represent results from \citet{Owen2016}, for which boil-off was performed at $900$K with initial atmospheric mass fractions of $0.3$ and $0.1$ in dot-dashed and dotted lines, respectively (note that initial atmospheric mass fractions in our models are calculated self-consistently through gaseous accretion). The right-hand panel shows planet sizes, including those that have been stripped of their atmosphere (atmospheric mass fraction $< 10^{-4}$) and sit on an Earth-like composition line (grey dashed). This panel can be interpreted as a primordial mass-radius distribution.}
    \label{fig:GridSummary} 
\end{center}
\end{figure*}

Figure \ref{fig:mass_grid_constT} shows a small parameter study for planets with core masses of $2M_\oplus$, $4M_\oplus$, $6M_\oplus$ and $8M_\oplus$ at temperatures of $600$K, $900$K, $1200$K, $1500$K. Under our assumptions of dust-free gas, planets further out in the disc will be at lower equilibrium temperatures and can, therefore, cool more efficiently, allowing accretion to occur at a quicker rate. In the same vein, a boil-off process will be less dramatic at lower equilibrium temperatures since it is the local irradiation that actually removes the material from the planet's influence (note the dependence of sound speed and thus the temperature in outflow velocity from Equation \ref{eq:D(r)}). However, accretion and boil-off are also strongly controlled by the core mass, since this sets the gravitational potential and, thus, the ease with which mass can enter/leave the planet. A clear trend can be seen in Figure \ref{fig:mass_grid_constT} in which planets with cooler equilibrium temperatures accrete a more massive atmosphere. Similarly, planets with larger mass cores follow the same trend. Indeed, the $8M_\oplus$ model at $600$ K undergoes runaway accretion before disc dispersal onset, terminating the evolutionary model, and is not shown in Figure \ref{fig:mass_grid_constT}.

% The evolution of planets in a pure boil-off scenario (no core luminosity and without considering XUV penetration) is shown with thin lines in Figure \ref{fig:mass_grid_constT}. We find that smaller core masses at larger equilibrium temperatures can be stripped (defined as an atmospheric mass fraction $X < 10^{-4}$) of their accreted H/He atmosphere within $\sim 1$ Myr of disc dispersal onset. 

In Figure \ref{fig:mass_grid_constT}, we demonstrate the transition from boil-off driven mass-loss (solid lines) to core-powered mass-loss (dashed lines, with transition denoted with squares) and XUV-driven photoevaporation (grey dotted, with transition denoted with circles). Figure \ref{fig:mass_grid_constT} demonstrates that boil-off can perform mass-loss at early ages for all considered masses and equilibrium temperatures. As discussed in Section \ref{sec:BO-CPML}, the energy/work done in driving atmospheric mass-loss during this boil-off phase (solid lines) is dominated by the release of binding energy due to atmospheric contraction and not contributions from the core. 

Recalling that the vast majority of mass-loss under boil-off/core-powered mass-loss occurs while the protoplanetary disc is present and optically thick, these results suggest that some small-mass, close-in planets may have been stripped before the disc has completely dispersed, independently of XUV photoevaporation. We discuss this point in Section \ref{sec:EvolutionBeyondboilOff}.

We present the properties of planet models from Figure \ref{fig:mass_grid_constT} at the onset of XUV photoevaporation in Figure \ref{fig:GridSummary}. Since these models adopt a disc dispersal timescale of $t_\text{disp}=3$~Myrs and XUV photons penetrate beneath the Bondi sphere $\sim$~Myr later, these conditions represent the planets at $\sim 4$~Myrs. Note that the exact time of XUV photoevaporative dominance would change for different disc ages and dispersal timescales. In the left-hand panel, we show atmospheric mass fractions compared to the results of \citet{Owen2016} in green squares. We discuss the differences in Section \ref{sec:breezes}; however, we highlight that the models of \citet{Owen2016} adopt different initial atmospheric mass fractions as a result of their different assumptions, many of which we have individually addressed in this work. 

% Similarly, we show the cooling timescales at the onset of XUV photoevaporation for our models in the middle panel of Figure \ref{fig:GridSummary}, given by Equation \ref{eq:Cooltimescale}. As highlighted in Figure \ref{fig:LcoreEffects}, the core increases the radiative luminosity of the planet while the atmospheric mass is decreasing, implying a shorter cooling timescale. In the absence of core luminosity and XUV-driven mass-loss (a pure boil-off scenario), we corroborate the results of \citet{Owen2016} in that planets have prematurely cooled such that their cooling timescales are $\sim 100$ Myr at $10$ Myr, as shown in Figure \ref{fig:5Mearth900K_VariousDiscDisp}. Note that the cooling timescale for planets with significantly small atmospheric masses e.g. $X \lesssim 10^{-3}$ are not shown in Figure \ref{fig:GridSummary} since these atmospheres have become completely radiative, yielding an extremely long cooling timescale.

In the right-hand panel of Figure \ref{fig:GridSummary}, we show the photospheric radius of each planet at the onset of XUV photoevaporation. For planets stripped of their hydrogen-dominated atmosphere, their photospheric radius equals their core radius, as highlighted with the mass-radius relation for Earth-composition cores from \citet{Fortney2007} in grey dashed. Since contributions from the core dominate planet mass, this panel can be interpreted as a mass-radius distribution for young planets at the end of disc dispersal and the onset of XUV photoevaporation. Planets at lower equilibrium temperatures are larger in size since they host more atmosphere post disc-dispersal. Note, however, that the exact size of a planet at a given time will depend on when the disc disperses and the rate at which this occurs. 

\subsection{Physical Disc Model} \label{sec:PhysicalDiscModels}

The final development of our evolution model is to consider a more realistic disc evolution sequence with a varying planet equilibrium temperature. Assuming that our newly formed planet sits at the midplane of the disc, there are three primary heat sources to consider: viscous heating due to stellar accretion, irradiation from dust in the upper disc atmosphere and direct stellar irradiation once the disc is optically thin through its midplane. Firstly, the flux released at the disc midplane due to active accretion \citep[e.g.][]{Cassen1994,Pringle1981} is:
\begin{equation} \label{eq:Fvisc}
    F_\text{visc} = \frac{3GM_* \dot{M}_*}{8 \pi a^3} \, \tau.
\end{equation}
where the factor of disc vertical optical depth $\tau = \Sigma_\text{g} \kappa / 2$ accounts for the flux at the midplane. The stellar accretion rate $\dot{M}_*$ is related to the gas surface density $\Sigma_\text{g}$ via:\footnote{Typically, one sees an additional factor of $1 - \sqrt{\frac{R_*}{a}}$ in this expression and in Equation \ref{eq:Fvisc}, which would arise from the assumption of a zero-torque inner boundary condition. Whilst appropriate for accretion discs orbiting black holes, the inner disc edge is truncated by the star's magnetic field in protoplanetary discs.}
\begin{equation} \label{eq:ActiveDisc_Sigma}
    \nu \Sigma_\text{g} = \frac{\dot{M}_*}{3 \pi},
\end{equation}
where $\nu$ is the kinematic viscosity, parameterised by the standard dimensionless $\alpha$-viscosity parameter \citep{ShakuraSunyaev}:
\begin{equation} \label{eq:ActiveDisc_Tmid}
    \nu = \alpha \frac{c_\text{s}^2}{\Omega_\text{K}},
\end{equation}
where we assume $\alpha = 10^{-2}$, which is appropriate for the inner disc where the viscosity is likely driven by the magneto-rotational instability \citep[e.g.][]{Jankovic2021,Delage2023}.
Secondly, we follow the 2-layer model of \citet{Chiang2001} for heating at the disc midplane due to passive irradiation from dust in the upper disc. A radiative balance is found in which:
\begin{equation} \label{eq:Tdust}
    \frac{\phi}{2}(1 - e^{-\Sigma_\text{g} \kappa_\text{s} }) \bigg( \frac{R_*}{a} \bigg)^2 T_*^4 \sin \beta = (1 - e^{-\Sigma_\text{g} \kappa_\text{i}}) \, T_\text{i}^4
\end{equation}
where $T_*$ and $R_*$ are the stellar effective temperature and radius respectively, $\kappa_\text{s}$ is the opacity of the disc interior to irradiation from the disc surface, $\kappa_\text{i}$ is the opacity of the disc surface to irradiation from the disc interior, $T_\text{i}$ is the disc interior temperature, $\phi=0.5$ is the fraction of the stellar hemisphere that is seen by a dust grain and $\beta$ is the grazing angle due to irradiation incident on a flaring protoplanetary disc:
\begin{equation} \label{eq:grazing_angle}
    \beta = \sin^{-1} \bigg( \frac{4}{3\pi} \frac{R_*}{a} \bigg ) + \tan^{-1} \bigg ( \frac{d \log z_\text{irr}}{d \log a} \frac{z_\text{irr}}{a} \bigg) - \tan^{-1} \bigg( \frac{z_\text{irr}}{a} \bigg ),
\end{equation}
where $z_\text{irr}$ is the vertical height at which irradiation is absorbed in the disc. For simplicity, we assume $z_\text{irr} = 4 H$ \citep{Chiang1997}, where $H$ is the disc vertical scale height. In hydrostatic equilibrium:
\begin{equation} \label{eq:discHSE}
    \frac{z_\text{irr}}{a} = \frac{z_\text{irr}}{H} \frac{H}{a} = \frac{z_\text{irr}}{H} \sqrt{\frac{T_\text{i}}{T_\text{g}}} \sqrt{\frac{a}{R_*}},
\end{equation}
where $T_\text{g} \equiv GM_* \mu m_\text{H} / k_\text{B} R_*$. We follow the numerical prescription from \citet{Chiang2001} to solve Equations \ref{eq:Tdust}, \ref{eq:grazing_angle} and \ref{eq:discHSE} for $z_\text{irr}$ and $T_\text{i}$ simultaneously on a grid of semi-major axis. For consistency with the rest of our dust-free gas calculations in this work, we adopt values of $\kappa_\text{s} = 0.1 \; \text{cm}^2 \; \text{g}^{-1}$ and $\kappa_\text{i} = 0.001 \; \text{cm}^2 \; \text{g}^{-1}$, although we confirm that our results are insensitive to our choices in disc opacities. Instead, temperatures are predominantly controlled by the host stellar radius and temperature. Although minor, the flux due to this source of dust-driven irradiation is given by:
\begin{equation} \label{eq:Fdust}
    F_\text{dust} = \sigma T_\text{i}^4.
\end{equation}
The final source of heating is direct stellar irradiation once the disc is optically thin through its midplane, with flux given by:
\begin{equation} \label{eq:Fstar}
    F_* = \frac{L_*}{4\pi a^2} e^{-\tau_\text{mid}},
\end{equation}
where $L_*$ is the stellar luminosity and $\tau_\text{mid}$ is the optical depth through the midplane of the disc:
\begin{equation} \label{eq:tau_discmid}
    \tau_\text{mid} = \int_{R_\text{in}}^a \rho_\text{mid}(r) \, \kappa_\text{disc} \, \mathrm{d}r.
\end{equation}
where we assume the disc's inner edge is truncated at $R_\text{in}=10R_*$ and the density at the disc midplane $\rho_\text{mid}(r)$ is related to the disc surface density via Equation \ref{eq:rhodisc}. To account for the pre-main-sequence (PMS) evolution of stellar luminosity $L_*$, effective temperature $T_*$ and radius $R_*$, we incorporate the \verb|MIST| evolution tracks for a $1 M_\odot$ star from \citet{MIST-I2016,MIST-II2016}. The equilibrium temperature $T_\text{eq}$ as a result of all three heating terms is thus given by:
\begin{equation} \label{eq:Teq_discEvolution}
    \sigma T_\text{eq}^4 = F_\text{visc} + F_\text{dust} + F_*, 
\end{equation}
where $F_\text{visc}$, $F_\text{dust}$, $F_*$ are given by Equations \ref{eq:Fvisc}, \ref{eq:Fdust} and \ref{eq:Fstar} respectively. To determine the evolution of the disc surface density and planet equilibrium temperature, we take the models of XUV-driven disc dispersal from \citet{Owen2011}. This dispersal mechanism relies on the trade-off between gas accretion onto the central star and gas being driven away vertically via photoevaporative winds generated by the high-energy flux from the star. As disc mass and accretion rate drop towards the end of the disc lifetime, there will come a point where gas is more likely to be removed by the disc wind than accreted. As a result, a gap is opened at this radius, typically $\sim 1$AU, which disconnects an inner and outer disc - causing the two to evolve separately. Since accretion discs evolve on a viscous timescale which scales linearly with disc radius,\footnote{For a constant $\alpha$, the kinematic viscosity $\nu \sim r$ and hence the viscous timescale $t_\text{visc} \sim \frac{r^2}{\nu}$ scales linearly with radius.} the inner disc rapidly drains onto the host star on timescales $\sim 10^5$ yrs, which triggers boil-off. 

In order to evolve the disc surface density, we assume an initial stellar accretion rate of $5 \times 10^{-8} M_\odot \text{ yr}^{-1}$ and use this to calculate an initial gas surface density profile $\Sigma(t=0, r)$ from Equation \ref{eq:ActiveDisc_Sigma}. We then normalise the models of disc evolution and dispersal from \citet{Owen2011} to have initial values of $\Sigma(t=0, r)$, which then provide $\Sigma(t, r)$ for all times and stellar separations. The resulting disc surface densities are used to calculate a planet's equilibrium temperature from Equation \ref{eq:Teq_discEvolution}. As a demonstration of this disc model, we calculate evolution for a planet at an orbital separation of $0.1$~AU and set the \verb|MESA| outer boundary temperature $T_\text{BC}$ (see Section \ref{sec:MESA_BC}) to these values as a function of time. 

\begin{figure} 
\begin{center}
	\includegraphics[width=0.95\columnwidth]{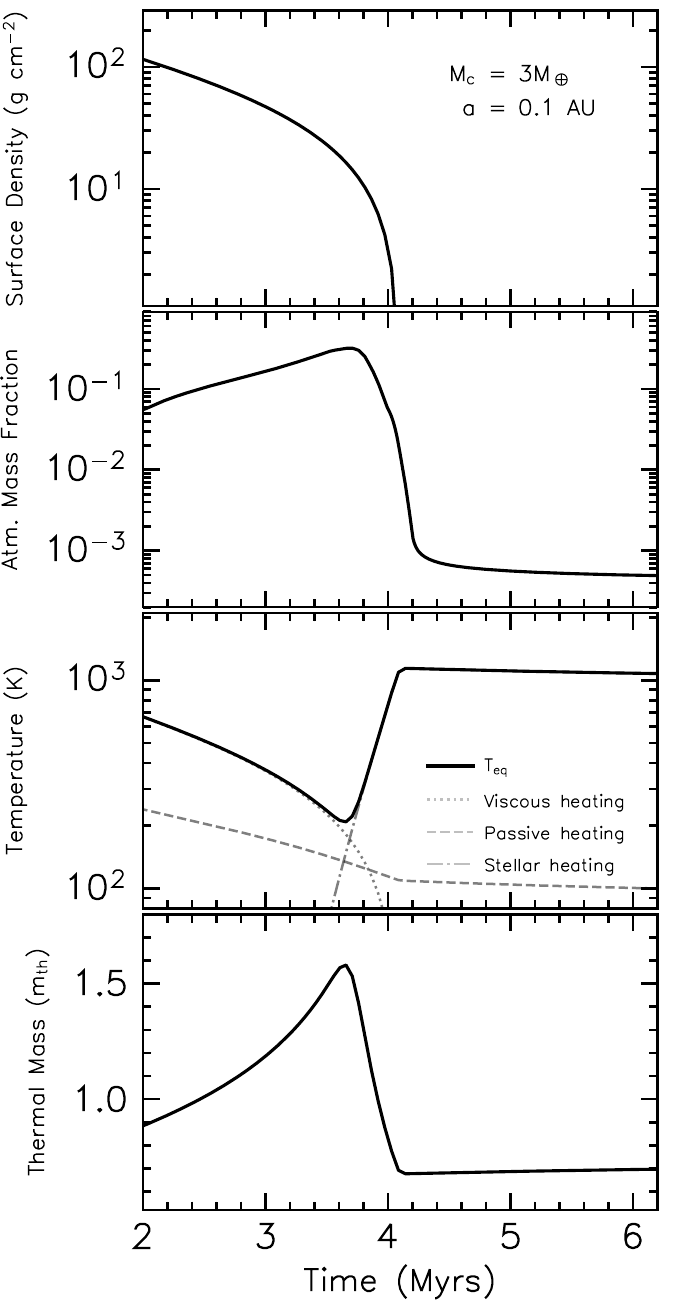}
    \caption{Evolution of disc surface density, atmospheric mass fraction, equilibrium temperature and thermal mass for a $3M_\oplus$ planet at $0.1$ AU. In the third panel, we show the contributions of planet equilibrium temperature from viscous heating (dotted), passive heating from dust in the upper protoplanetary disc (dashed), and direct stellar heating which dominates once the disc is optically thin through its midplane (dot-dashed). In the bottom panel, the thermal mass represents the ratio of the planet's Hill radius to the disc's vertical scale height. If this ratio rises above unity, it indicates that a gap may open in the protoplanetary disc.}
    \label{fig:DiscModel} 
\end{center}
\end{figure}

In Figure \ref{fig:DiscModel}, we plot the disc surface density, atmospheric mass fraction, and equilibrium temperature with contributions from viscous, dust and stellar heating in dotted, dashed and dot-dashed lines respectively. We show this as a function of time, as opposed to time since disc dispersal as in Figures \ref{fig:5Mearth900K_Entropy} - \ref{fig:mass_grid_constT} since the disc dispersal starts gradually and without a distinct onset in these more realistic disc evolution models. As in our previous simpler disc models from Sections \ref{sec:tcool_vs_tdisp} and \ref{sec:BO-CPML-XUV}, the planet accretes gas and then undergoes a rapid mass-loss phase as the disc disperses. One can see that the equilibrium temperature initially falls at $\sim2$~Myr since the disc is optically thick and the surface density, accretion rate and therefore viscous heating are all dropping. In addition, PMS stellar evolution at this stage also contributes to a decreasing level of stellar irradiation, which reduces heating due to dust. The disc becomes optically thin at $\sim 4$ Myr, revealing the planet to direct stellar irradiation for the first time. The equilibrium temperature rapidly rises and is dominated by the stellar luminosity. 

In the bottom panel of Figure \ref{fig:DiscModel}, we show the evolution of the planet's dimensionless thermal mass $m_\text{th}$ \citep[e.g.][]{LinPapaloizou1985}, defined as the ratio between a planet's Hill radius and its nascent disc's vertical scale height:
\begin{equation}
    m_\text{th} = a \sqrt[3]{\frac{M_\text{p}}{3 M_*}} \frac{\Omega_\text{K}}{c_\text{s}}.
\end{equation}
The importance of thermal mass relates to the ability of a planet to open a gap in the protoplanetary disc. If $m_\text{th} > 1$, then the Hill sphere extends beyond the disc scale height and implies sufficient local gravitational influence that gap formation might occur. Figure \ref{fig:DiscModel} implies that gap opening could occur, even for a small $3M_\oplus$ planet. The reason is due to the strong dependence of thermal mass on disc temperature, which drops considerably before the host star directly irradiates the planet. Rapid gap opening may induce enhanced boil-off, as atmospheric mass is lost to the void opened in the planet's vicinity. We discuss this in Section \ref{sec:coldsunrise}.

\section{Discussion} \label{sec:discussion}

We have presented new evolutionary models of small, close-in exoplanets undergoing gaseous accretion, boil-off and core-powered mass-loss. They highlight the importance of disc dispersal in setting the initial conditions for planetary evolution in the post-disc epoch. They also demonstrate the need for self-consistent modelling, from gas accretion to direct irradiation from the host star. We have shown that planets will typically accrete a few to tens of per cent in atmospheric mass before boil-off begins, although we highlight that the exact values depend on parameters such as core mass, orbital separation, disc evolution, gas opacities and the time at which gaseous accretion began. Furthermore, once the disc begins to disperse, boil-off is triggered and removes $\gtrsim 90\%$ of atmospheric mass, again with exact values depending on core mass, orbital separation and gas opacities. Crucially, the ratio of cooling timescale to disc dispersal timescale is imprinted on the final atmospheric mass fraction of the planet, with quicker dispersal times inducing a more significant change in atmospheric mass for a given planet's cooling timescale. In addition, we have considered the role of core luminosity and the transition from boil-off, to core-powered mass-loss and XUV-driven photoevaporation. We now discuss the implications of these findings. 

% {\jr{Whilst we have only considered accretion in a dust-free environment, we note that its presence may also not considered gas accretion in the presence of dust, which is known to increase gas opacity and thereby slow accretion rates \citep[e.g.][]{Ikoma2000,Piso2015}, this result is consistent with previous works that have considered accretion in the dust-free case within their analyses \citep[e.g.][]{Ikomi2012,Lee2014,Lee2015,Ginzburg2016,Lee2018,FungLee2018}.}} 

\subsection{A Consistent Evolution Sequence} \label{sec:PEvsAcc}

{\jr{In \citet{Jankovic2018}, planet formation in the inner protoplanetary disc was studied for small-mass exoplanets. They modelled dust evolution, core accretion and then XUV-driven photoevaporation of planet atmospheres following disc dispersal, but excluding boil-off. The final mass and radii of these planets were shown to be inconsistent with current observational data, with the models producing a higher atmospheric mass fraction than is observed. Then, in \citet{Rogers2021} and \citet{Rogers2023}, XUV photoevaporation was exploited to rewind the clock for \textit{Kepler} planet evolution. To be consistent with the present-day demographics, they showed that the initial atmospheric mass fraction of planets, once their discs had dispersed, must scale positively with core mass, consistent with the analytic findings of \citet{Ginzburg2016}. When considering the entire exoplanet population with a peaked core mass distribution, this scaling results in typical sub-Neptune initial atmospheric mass fractions of $\sim 2\%$ \citep[see also][]{Wu2019}. On the other hand, many studies considering accretion under a range of conditions have shown that in the dust-free case, as considered in this work, planets $\gtrsim 5M_\oplus$ can accrete $\gtrsim 10 \%$ in atmospheric mass by the end of the disc lifetime \citep[e.g.][]{Ikomi2012,Lee2014,Lee2015,Ginzburg2016,Lee2018,FungLee2018}, which is in tension with the value of $\sim 2\%$ found above. Note that this discrepancy is likely not an indication that XUV photoevaporation underestimates mass-loss rates since planets beyond orbital periods of $\sim 30$ days (which are not strongly affected by photoevaporation) have mass and radius measurements consistent with hosting a few per cent in H/He mass, instead of $\gtrsim 10\%$ \citep[e.g.][]{Weiss2014,JontofHutter2016,Wolfgang2016}. As we have shown in this work, boil-off can provide a potential solution to this discrepancy via significant atmospheric escape during disc dispersal. We also highlight that, in contrast to some of the photoevaporation models discussed above, core-powered mass-loss models are typically initialised with the atmospheric mass fraction scaling of \citet{Ginzburg2016}, thus implicitly assuming boil-off in their initial conditions \citep{Ginzburg2018,Gupta2019,Gupta2020}. As such, the aforementioned discrepancy was not seen in their work.

% In this work, we have shown that atmospheric escape, induced by the rapid dispersal of the local protoplanetary disc, can resolve such tensions. Whilst planets are shown to accrete $\gtrsim 10\%$ in atmospheric mass, they rapidly lose this mass during disc dispersal to a distribution centred at $\sim 1\%$, which is consistent with the results from \citet{Rogers2021}. Note that this latter study also inferred the core mass distribution of small exoplanets to find a peak of maximum occurrence $\sim 4M_\oplus$. As such, there are relatively few planets with core masses $\gtrsim 10M_\oplus$, for which we have shown are less vulnerable to boil-off (see Figure \ref{fig:GridSummary}). 

Whilst we have shown boil-off is an important process and may provide a solution to the aforementioned discrepancy, other mechanisms do exist that may also resolve this tension. For example, one important consideration that has not been included in this work is the effects of gas opacity on accretion \citep[e.g.][]{Ikoma2000,Piso2015}. As additionally considered in multiple studies \citep{Ikomi2012,Lee2014,Lee2015,LeeChiang2016,Lee2018,FungLee2018}, dust acts to increase gas opacity in the planet's atmosphere and surrounding protoplanetary disc, which thus slows cooling and the rate of gas accretion. In this case, typical accreted atmospheres are $\lesssim 10\%$, thus reducing the significance of the discrepancy between accretion and mass loss. Additional ways to resolve tensions include: forming the planet at the very end of the disc lifetime \citep{LeeChiang2016}; accretion-induced shocks \citep{Lee2019}, increased planetary luminosity as a result of giant mergers \citep{Liu2015,Inamdar2016}; or recycling of high-entropy material from the disc, although we highlight that the efficacy of this mechanism is currently under debate \citep{Ormel2015,Fung2015,Cimerman2017,Kurokawa2018,AliDib2020,Chen2020,Bailey2023,Savignac2023}. Indeed, we find that, in the lack of such mechanisms, many planets undergo runaway accretion before disc dispersal, particularly in the case of more realistic disc evolution models. Since we observe many more sub-Neptunes with atmospheric mass fractions $\lesssim 5 \%$ compared to Jovian planets, there must exist mechanisms such as those described that slow or delay the rate of accretion prior to disc dispersal.}}

\subsection{A Gentle Breeze} \label{sec:breezes}
Previous studies that considered mass-loss during disc dispersal have taken different approaches, yielding different results. In \citet{Ikomi2012}, accretion and mass-loss were considered in a hydrostatic equilibrium regime. They found that some planets will continue to lose their entire atmospheric masses as the disc disperses to maintain this equilibrium. \citet{Owen2016}, on the other hand, argued that planets will not be able to stay in hydrostatic equilibrium with the disc since dispersal occurs on timescales of $10^5$ yrs \citep{Kenyon1995,Ercolano2011,Koepferl2013}. In contrast, the cooling timescale of a planet has upper limits of $t_\text{cool} \sim 10^6 - 10^7$ yrs. In the case that hydrostatic equilibrium cannot be maintained, mass-loss will ensue via a hydrodynamic wind. \citet{Owen2016} assumed the planet was initially fully convective out to the Bondi radius, with an escaping isothermal Parker wind \citep{Parker1958} and an instantaneous disc dispersal. Unlike \citet{Ikomi2012}, they found that planets do not lose all their mass, since mass-loss shuts off once the planet shrinks to small sizes $\sim 0.1 R_\text{B}$. However, their assumptions implied that their calculated mass-loss rates were likely upper limits.

In this study, we have built upon the models of \citet{Ikomi2012} and \citet{Owen2016}, by self-consistently modelling the planet through gas accretion and disc dispersal. Notably, the planet is allowed to remain in causal contact with the disc via an isothermal outflow and non-hydrostatic equilibrium if this is not physically possible. When introducing this self-consistent approach, we find that the planets require physics from both \citet{Ikomi2012} and \citet{Owen2016}. Planets begin by cooling and accreting, before losing mass via subsonic breeze solutions, allowing the planet to sustain causal contact with the disc as it disperses. As this continues, the Mach number of the outflow must continually increase such that the escaping gas eventually becomes transonic (see Figure \ref{fig:5Mearth900K_Entropy}). We find that once this transition occurs, the planet leaves causal contact with the disc and, similar to \citet{Owen2016}, the mass-loss then quickly shuts off once the planet has shrunk to $\sim 0.1 R_\text{B}$. We highlight that our models suggest that the vast majority of mass-loss occurs while the disc survives. Therefore, boil-off/core-powered mass-loss has all but concluded by the time the disc has dispersed in most cases. As a result, boil-off is likely unaffected by the dynamical instabilities highlighted in \citet{Wang2023} due to mass-loss in multi-planet systems. During periods of high mass-loss rate, the disc is present and likely dissipates any orbital instabilities that arise due to changes in angular momentum. Although, we highlight that more work should be done to investigate this.

We find that the majority of boil-off cooling occurs due to advection \citep[as initially shown in][]{Owen2016} and enhanced radiative cooling as atmospheric mass is rapidly lost, inducing steep temperature gradients due to $P \textrm{d} V$ expansion. We additionally corroborate the findings of \citet{Owen2016} in that the typical cooling timescales for planets that have undergone a boil-off are $\tau_\text{cool} \sim 10^8$ yrs at $10$ Myrs (see Figures \ref{fig:5Mearth900K_VariousDiscDisp} and \ref{fig:LcoreEffects}). Thermodynamically speaking, planets appear older than their age would suggest due to significant premature cooling caused by mass-loss during boil-off. As shown in \citet{Owen2020}, this means that observations of young planets with measured masses, radii and ages may allow us to constrain their initial entropies and, therefore, determine whether they did or did not undergo a boil-off phase.

\subsection{Boil-off sets the initial conditions for planetary evolution} \label{sec:formationhistory}
One of the key findings from this study has been the impact of disc dispersal timescale on atmospheric evolution. For a given planet mass and equilibrium temperature, the final state of a planet after disc dispersal is predominantly controlled by the ratio of a planet's cooling timescale to disc dispersal timescale. As we have shown in Figure \ref{fig:5Mearth900K_Entropy}, larger ratios of cooling timescale to disc dispersal timescale $\tau_\text{cool} / \tau_\text{disc}$ induce greater levels of advection and radiative cooling due to $P \textrm{d} V$ expansion. This then causes planet evolution to approximately converge, independently of initial conditions. Nevertheless, very large ratios lead to a small dispersion in the final atmospheric mass (see the middle-left panel of Figure \ref{fig:5Mearth900K_VariousDiscDisp}). This is indeed the conclusion that \citet{Owen2016} came to under their assumption of an instantaneous disc dispersal, suggesting that our models are consistent with theirs in the limit of $\tau_\text{cool} / \tau_\text{disc} \rightarrow \infty$.

In the more realistic case of a non-instantaneous disc dispersal based on observed timescales of $\sim 10^5$ yrs \citep{Kenyon1995,Ercolano2011,Koepferl2013}, our results suggest that the entropy, and therefore thermal structure of the planet, is set by boil-off. Information about the ratio of a planet's cooling timescale to disc dispersal timescale is imprinted on the atmospheric structure of the planet, which will affect how a planet evolves after the disc has dispersed. This opens up the possibility of using demographic surveys of exoplanets to perform physics-based inference with boil-off, core-powered mass-loss and XUV photoevaporation models to place constraints on disc dispersal timescales, which is left for future work.

\subsection{Comparison with Ginzburg et al. (2016)}

In \citet{Ginzburg2016}, a purely analytic approach, based on radiative-convective equilibrium models of H/He atmospheres, was used to predict the atmospheric mass $X \equiv M_\text{atm} / M_\text{core}$ at the end of the boil-off phase (referred to as `spontaneous mass-loss' in their study). Unlike \citet{Owen2016}, their initial conditions were analytically calculated self-consistently with gaseous accretion prior to disc dispersal. Under the assumption of constant gas opacity, ideal gas equations of state and an adiabatic index of $\gamma = 7/5$, they predict:
\begin{equation} \label{eq:Ginzburg2016}
    X_\text{final} \approx 0.01 \, \bigg( \frac{M_\text{c}}{M_\oplus} \bigg)^{0.44}  \bigg( \frac{T_\text{eq}}{1000 \text{ K}} \bigg)^{0.25} \bigg( \frac{t_\text{disp}}{1 \text{ Myr}} \bigg)^{0.50}.
\end{equation}
As shown in Figure \ref{fig:GridSummary}, we find a steeper dependence of atmospheric mass fraction with core mass after boil-off and an inverse dependence on equilibrium temperature. In other words, we find that hotter planets lose more atmosphere. This latter difference arises from our more realistic modelling of opacities from \citet{Freedman2008}, which vary with temperature. Indeed, if one performs the analysis of \citet{Ginzburg2016} with temperature-dependant opacities, an inverse dependence of atmospheric mass fraction with equilibrium temperature can be recovered. We speculate that the steeper mass-scaling found in this work compared to \citet{Ginzburg2016} and inferred in \citet{Rogers2023} is likely due to the fact that we are unable to model the gas-accretion phase on a population level over enough cooling timescales before the onset of disc dispersal. We leave the details of understanding these different mass scalings for future work.

% Using energy arguments alone, \citet{Ginzburg2016} showed that boil-off causes planets to converge to a radius for which $R_\text{p} / R_\text{core} \approx 2$ \citep[this boundary separates `thick' and `thin' atmospheric regimes in][]{Ginzburg2016}. The assumption is used in the derivation of Equation \ref{eq:Ginzburg2016} and is also frequently used as initial conditions for core-powered mass-loss models \citep[e.g.][]{Ginzburg2018,Gupta2019,Gupta2020}. In our new boil-off models, however, we find that the $R_\text{p} / R_\text{core}$ is not constant but instead increases for larger core masses and lower equilibrium temperatures (see Figure \ref{fig:GridSummary}). This may explain the weak dependence on core mass found in Equation \ref{eq:Ginzburg2016}, since a constant $R_\text{p} / R_\text{core}$ will only scale with $R_\text{core} \propto M_\text{core}^{1/4}$ \citep{Valencia2006}. Our self-consistent models show that the atmospheric mass fraction at the end of boil-off scales approximately as $M_c^2$.

\subsection{Evolution beyond boil-off} \label{sec:EvolutionBeyondboilOff}

% \begin{figure*} 
% \begin{center}
%     \includegraphics[width=2.0\columnwidth]{Figures/PopulationEvolution.pdf}
%     \caption{Synthetic populations of planets evaluated with the semi-analytic model of \citet{Owen2017,Rogers2021}. Left-hand panel: we take atmospheric mass fractions and Kelvin-Helmholtz timescales from the end of our boil-off calculations (see Figures \ref{fig:GridSummary}), combined with the core mass distribution from \citet{Rogers2021} to yield a planet population evaluated at $10$~Myr. Observations (shown in black) are taken from the NASA exoplanet archive for stellar ages $< 100$~Myr. Right-hand panel: The models are then evolved through cooling and XUV photoevaporative mass-loss evolution to $5$~Gyr. \textit{Kepler} observations in this panel are from \citet{Ho2023}. The colours in both panels represent each planet's core mass. We highlight that we do not correct models for observational uncertainty or biases, which strongly affect the data.}
%     \label{fig:PopulationEvolution} 
% \end{center}
% \end{figure*}

Once the initially rapid mass-loss, cooling and contraction involved in the boil-off phase have ceased, the atmospheric evolution of small, close-in planets will be dictated by atmospheric processes such as core-powered mass-loss \citep[e.g.][]{Ginzburg2018,Gupta2019} and/or XUV-photoevaporation \citep[e.g.][]{LopezFortney2013,Owen2013}. In Section \ref{sec:BO-CPML-XUV}, we considered the effect of core luminosity (arising from core cooling and radiogenic heating) and its influence on atmospheric evolution. As boil-off progresses and the core cooling luminosity increases, it will eventually dominate the planet's energy budget. As such, the core provides the energy required to lift gas to the base of the bolometrically heated outflow. This changeover offers a convenient definition for the transition between boil-off and core-powered mass-loss, as highlighted in Figure \ref{fig:LcoreEffects}. We have demonstrated in Figure \ref{fig:mass_grid_constT} that larger atmospheres around larger planet cores release more energy and transition to core-powered mass-loss later. Nevertheless, the bottleneck to boil-off or core-powered mass-loss is the speed at which bolometric heating can remove material via a Parker outflow. As such, the final atmospheric mass fractions with and without core luminosities are very similar, as shown in Figure \ref{fig:LcoreEffects}. 

As described in \citet{Owen2023b}, XUV photons from the host star will eventually penetrate beneath the sonic surface of the planet at $R_\text{B}$. In this case, XUV photons can heat and ionise the hydrogen-dominated gas before it reaches transonic velocities and becomes causally disconnected from the planet's interior. As such, a strong XUV-driven photoevaporative wind is launched from beneath the Bondi surface. This defines the transition from boil-off or core-powered mass-loss to XUV photoevaporation, which we highlight in Figure \ref{fig:mass_grid_constT}. One can see that some smaller-mass planets are affected by all three evolution processes: boil-off, core-powered mass-loss and photoevaporation. This agrees with the speculated evolutionary scenario proposed by \citet{Owen2023b}. The smallest, highly irradiated planets are stripped of their atmospheres during disc dispersal prior to XUV heating. We note that throughout this work, we have defined a stripped planet as one with a negligible atmospheric mass fraction $X < 10^{-4}$. As shown in \citet{Misener2021}, however, small-mass primordial atmospheres ($X \lesssim 10^{-5}$) can be retained in the scenario of core-powered mass-loss. Nevertheless, such small atmospheres are unlikely to be retained in the XUV photoevaporation regime unless one gets significant mass-dependent fractionation.

Our models suggest that many small-mass planets may have been stripped of their primordial atmospheres through a combination of boil-off and core-powered mass-loss. This creates a radius valley prior to the effects of photoevaporation, which typically acts on $\sim 100$~Myr to Gyr timescales. Note, however, that this is distinct from the `primordial' radius valley as predicted in \citet{Lee2021,Lee2022} which arises purely due to gas accretion and would therefore be present even before boil-off / core-powered mass-loss. Regardless, the combination of these proposed mechanisms suggests that the radius valley should be observable in the demographics of very young planetary systems.

\subsection{A Cold Sunrise} \label{sec:coldsunrise}
In Section \ref{sec:PhysicalDiscModels}, we considered a more realistic disc evolution model that included an evolving disc midplane temperature. As shown in Figure \ref{fig:DiscModel}, the equilibrium temperature experienced by a close-in planet drops, often considerably, below its future value once the disc is optically thin and the planet is directly irradiated by its star for the first time. The effect of this is two-fold: firstly, cooler temperatures in a dust-free environment increase gaseous accretion rates since planets can cool and contract more efficiently to accrete more material within their Bondi sphere. Secondly, cooler temperatures reduce the sound speed of the escaping gas, which reduces the mass-loss rate as the disc disperses. Nevertheless, we still find that planets accrete $\gtrsim 10\%$ in H/He dominated material and then boil-off during disc dispersal to atmospheric mass fractions of order $\sim 1\%$. As discussed in Section \ref{sec:PEvsAcc}, we also find that many planets with larger core masses undergo runaway gaseous accretion due to the lower temperatures. However, since we observe many planets of mass $\gtrsim 5M_\oplus$ with H/He atmospheric mass fractions $\lesssim 10\%$, there must exist mechanisms by which gaseous accretion is reduced or delayed to be consistent with results \citep[e.g.][]{Lee2019}. In Figure \ref{fig:DiscModel}, we also evaluated the planet's thermal mass to show that even a small $3M_\oplus$ planet may open a gap in the protoplanetary disc towards the end of disc dispersal. Under such circumstances, accretion can only occur through the gap, possibly via a circumplanetary disc, yielding lower mass fluxes.

This latter point raises the question of whether gap-opening, akin to the local gas disc dispersing on very quick timescales, can induce some form of mass-loss? Whilst we leave this for future work, one can consider larger-mass planets that have already opened gaps prior to boil-off. The act of opening a gap may induce mass-loss as the local gas density rapidly drops. {\jr{On the other hand, when the disc does disperse, as considered in this work, the planet will not experience any further boil-off since the planet has already been exposed to a vacuum in its local vicinity during the gap-opening process.}} Indeed, \citet{Rogers2023_waterworld} found tentative evidence for the lack of boil-off in evolved planets with measured mass and radii around M-dwarfs. An inferred increase in atmospheric mass fraction dispersion for planets $\gtrsim 10M_\oplus$ suggested that they did not follow the physics of boil-off, leading to far greater initial atmospheric masses $\sim 10\%$ after disc dispersal. This could be caused by slow disc dispersal rates, or gap-opening as discussed above.

\subsection{Model Improvements} \label{sec:boiloffImprovements}

We have shown that self-consistent modelling of gas accretion and mass-loss can shed light on the atmospheric evolution of exoplanets during disc dispersal. However, there is one major assumption that we have made that limits this analysis: an isothermal outflow. As shown in Figure \ref{fig:BoundaryCond}, the assumption of isothermality is likely valid in the radiative zone of a planet. However, this may not be the case, particularly in optically thin regions far from the planet. In assuming an isothermal outflow, we inherently assume that energy is inserted into the flow to keep it heated to $T_\text{eq}$ and thus remove it from the planet's Bondi/Hill sphere. In reality, there may be circumstances where the flow cannot be heated to sufficient temperatures to leave the planet's influence. As discussed in Section \ref{sec:breezes}, mass-loss in the current model shuts off due to the planet cooling and contracting. However, this effect may be compounded when considering non-isothermal outflows. One could use hydrodynamic models that consider the radiative transfer and subsequent mass-loss in the upper atmosphere to understand this effect. These models can then be used to construct a grid of mass-loss rates, which can be implemented in \verb|MESA|. This was indeed the approach of \citet{Owen2013}, applied to XUV photoevaporation, in using the mass-loss rates from hydrodynamic models from \citet{OwenJackson2012}. This is left for future work.

Finally, a significant effect that has been neglected is that of dust. Thus far, only dust-free gas has been considered by way of the opacity tables from \citet{Freedman2008} utilised in atmospheric structure, and our choices in disc opacities. In reality, dust is also present in the protoplanetary disc and, subsequently, the planet's atmosphere. Since opacity changes the rate of cooling and therefore the rate of accretion and atmospheric escape, it will likely alter the exact amount of mass that is accreted onto, or lost from, sub-Neptune atmospheres during disc dispersal \citep[e.g.][]{Owen2016}. Investigating this effect is also left for future work.

\section{Conclusion} \label{sec:conclusion}
We have considered the atmospheric evolution of small, close-in exoplanets hosting H/He-dominated atmospheres through protoplanetary disc dispersal. As the surrounding disc gas dissipates, a powerful hydrodynamic outflow is triggered from the planet, inducing rapid cooling and contraction of the atmosphere. We have shown that this process, referred to as `boil-off', is instrumental in setting the initial conditions for other mass-loss mechanisms, such as core-powered mass-loss and XUV photoevaporation. Our main conclusions are as follows:
\begin{itemize}

    \item {\jr{Planets $\lesssim 10M_\oplus$ will typically accrete a few to tens of per cent in atmospheric mass during the disc lifetime, depending on disc conditions.}} The onset of disc dispersal then triggers boil-off, in which planets typically lose $\gtrsim 90\%$ of their accreted atmospheres. This mechanism provides a possible solution to the documented discrepancy between predicted accreted masses and the initial conditions inferred from \textit{Kepler} data with XUV photoevaporation, as shown in \citet{Jankovic2018,Rogers2021}.
    \newline
    
    \item Unlike previous studies of mass-loss during disc dispersal \citep[e.g.][]{Ikomi2012,Owen2016}, we show that boil-off occurs via subsonic breeze outflows, which allow a causal connection to exist between escaping atmosphere and disc. As such, boil-off has all but concluded by the time the disc has fully dispersed. This typically coincides with the planet launching a transonic wind, which yields relatively low mass-loss rates due to the planet's contracted size.
    \newline
    
    \item For a given planet mass and equilibrium temperature, we find that the amount of mass lost during boil-off is directly dictated by the ratio of cooling timescale to disc dispersal timescale. From disc observations, one expects this ratio to be greater than unity; however, the larger the ratio, the more mass is lost. This leads to convergent evolution, largely independent of how much atmosphere was initially accreted. We show that this is due to advective cooling and a brief period of $P \mathrm{d}V$ expansion with subsequent radiative cooling that occurs at the onset of boil-off.
    \newline

    % \item We have shown that boil-off is capable of stripping small mass ($\lesssim 2M_\oplus$), close-in ($T_\text{eq} \gtrsim 1000$~K) planets of their primordial atmospheres even before the disc has fully dispersed, without the need of late-time mass-loss processes such as core-powered mass-loss or XUV photoevaporation. This implies the existence of a population-level radius gap existing prior to the latter escape processes taking effect.
    % \newline

    \item The effect of core luminosity, with contributions from core cooling and radioactivity, can eventually dominate a planet's energy budget once atmospheric contraction has slowed, marking the transition to core-powered mass-loss \cite[e.g.][]{Ginzburg2016,Gupta2019,Misener2021}. We have shown that a combination of boil-off and core-powered mass-loss is capable of stripping small mass ($\lesssim 2M_\oplus$), close-in ($T_\text{eq} \gtrsim 1000$~K) planets of their primordial atmospheres even before the disc has fully dispersed, without the need of late-time mass-loss processes such XUV photoevaporation.
    \newline

    \item Eventually, XUV photons can penetrate below a planet's sonic surface and induce a more powerful photoevaporative hydrodynamic escape. We follow the work of \citet{Owen2023b} and corroborate that the radii of smaller mass planets are predominantly sculpted by core-powered mass-loss. In contrast, higher-mass planets are sculpted by photoevaporation.
    \newline

    \item Finally, we considered a more realistic disc evolution model with evolving midplane temperature to show that planets may open gaps in their protoplanetary disc during boil-off. The gap may enhance mass-loss for low-mass planets due to rapid reduction in local confining pressure. For high mass planets $\gtrsim 10M_\oplus$, this may occur during its accretion phase and limit the amount of gas that can flow onto the planet. More work is needed to understand this process.    
\end{itemize}

We highlight that the assumptions of an isothermal outflow and dust-free opacities have limited this study. In future work, we plan to incorporate such effects to understand the radiative processes that limit/enhance mass-loss during disc dispersal. 

\section*{Acknowledgements}
We would like to kindly thank the anonymous reviewer, as well as Richard Booth, Akash Gupta, Eve Lee, Will Misener, Subu Mohanty, Ruth Murray-Clay, Matthäus Schulik and Yao Tang for discussion and comments that helped improve this work. JGR is sponsored by the National Aeronautics and Space Administration (NASA) through a contract with Oak Ridge Associated Universities (ORAU). JGR was also supported by the Alfred P. Sloan Foundation under grant G202114194 as part of the AEThER collaboration. JEO is supported by a Royal Society University Research Fellowship. JEO has also received funding from the European Research Council (ERC) under the European Union's Horizon 2020 research and innovation programme (Grant agreement No. 853022, PEVAP). H.E.S. gratefully acknowledges NASA grant 80NSSC18K0828 for financial support during preparation and submission of the work. For the purpose of open access, the authors have applied a Creative Commons Attribution (CC-BY) licence to any Author Accepted Manuscript version arising.

%%%%%%%%%%%%%%%%%%%%%%%%%%%%%%%%%%%%%%%%%%%%%%%%%%
\section*{Data Availability}

The models presented in this work are available from the authors upon reasonable request.

%%%%%%%%%%%%%%%%%%%% REFERENCES %%%%%%%%%%%%%%%%%%

% The best way to enter references is to use BibTeX:

\bibliographystyle{mnras}
\bibliography{references} % if your bibtex file is called example.bib

% Alternatively you could enter them by hand, like this:
% This method is tedious and prone to error if you have lots of references
%\begin{thebibliography}{99}
%\bibitem[\protect\citeauthoryear{Author}{2012}]{Author2012}
%Author A.~N., 2013, Journal of Improbable Astronomy, 1, 1
%\bibitem[\protect\citeauthoryear{Others}{2013}]{Others2013}
%Others S., 2012, Journal of Interesting Stuff, 17, 198
%\end{thebibliography}

%%%%%%%%%%%%%%%%%%%%%%%%%%%%%%%%%%%%%%%%%%%%%%%%%%

%%%%%%%%%%%%%%%%% APPENDICES %%%%%%%%%%%%%%%%%%%%%

% \appendix
% \section{Some extra material}

%%%%%%%%%%%%%%%%%%%%%%%%%%%%%%%%%%%%%%%%%%%%%%%%%%

% Don't change these lines
\bsp	% typesetting comment
\label{lastpage}
\end{document}